\pdfoutput=1
\documentclass[10pt,conference,compsocconf]{IEEEtran}
\usepackage{amssymb}
\usepackage{amsmath}
\usepackage{hyperref}
\usepackage{cleveref}
\usepackage{subcaption}
\usepackage{tikz}
\usepackage{multirow}
\usepackage{flushend}

\def\code#1{\texttt{#1}}
\def\subheading#1{\noindent{\textbf{#1: }}}

\newcommand{\ignore}[1]{}
\newcommand{\system}{MATRIX}
\newcommand{\privo}{PrivoScope}

\begin{document}

\title{Mitigating Location Privacy Attacks on Mobile Devices using Dynamic App Sandboxing}
\author{
\IEEEauthorblockN{Sashank Narain}
\IEEEauthorblockA
	{College of Computer and Information Science\\
	Northeastern University, Boston, MA, USA\\
	Email: sashank@ccs.neu.edu}
\and
\IEEEauthorblockN{Guevara Noubir}
\IEEEauthorblockA
	{College of Computer and Information Science\\
	Northeastern University, Boston, MA, USA\\
	Email: noubir@ccs.neu.edu}
}
\maketitle

\begin{abstract}
  We present the design, implementation and evaluation of a system, called
  \system, developed to protect the privacy of mobile device users
  from location inference and sensor side-channel
  attacks. \system~gives users control and visibility over 
  location and sensor (e.g., Accelerometers and Gyroscopes) accesses by mobile
  apps. It implements a \emph{PrivoScope} service that audits all location
  and sensor accesses by apps on the device and generates
  real-time notifications and graphs for visualizing these accesses;
  and a \emph{Synthetic Location} service to enable users to provide
  obfuscated or synthetic location trajectories or sensor traces to
  apps they find useful, but do not trust with their private
  information. The services are designed to be extensible and easy for
  users, hiding all of the underlying complexity from
  them. \system~also implements a \emph{Location Provider} component
  that generates realistic privacy-preserving synthetic identities and
  trajectories for users by incorporating traffic information using 
  historical data from Google Maps Directions
  API, and accelerations using statistical information from user driving experiments. 
  The random traffic patterns are
  generated by modeling/solving user schedule using a randomized
  linear program and modeling/solving for user driving behavior using a
  quadratic program. We extensively evaluated MATRIX using user studies, 
  popular location-driven apps and machine learning techniques, and 
  demonstrate that it is portable to most Android devices globally, 
  is reliable, has low-overhead, and generates synthetic 
  trajectories that are difficult to differentiate from real mobility 
  trajectories by an adversary.
\end{abstract}

\section{Introduction}

Mobile smartphones are presently the primary means for users globally to communicate, access information and even interact with the physical environment. These devices are equipped with an increasingly large number of precise and sophisticated sensors. These sensors vastly improve the quality of the user's interaction with the environment, but also pose significant threats for privacy breaches as they directly or indirectly leak private information about their users. The leakage of location information from the GPS sensor, for instance, has been a fast growing privacy concern. The commercial GPS hardware available in modern smartphones is capable of triangulating a user's position within an accuracy of 3 meters. This leakage enables more sophisticated threats such as tracking users, identity discovery, and identification of home and work locations. 

\subheading{Motivation} 
The current protections against location tracking mostly revolve around obfuscating the users' location. Several research works have proposed solutions that induce noise in the location data~\cite{Andres2013, Bordenabe2014, Shokri2012, Wang2012, ardagna2011}. Others have devised solutions that sends the real location with several dummy locations or within a data-set, and uses the response pertaining to the real location~\cite{Hoh07, Kato2012, Kido2005, Lu2008, Suzuki2010, You2007}. Others have proposed stripping off all identifying information about a user before sending the real location data in order to protect the user's privacy~\cite{pingley2011, Baik2006}. Unfortunately, these solutions still leak information about their users and can be combined with other information (e.g., census data) to infer user identities and their locations~\cite{Shokri2011, Zang2011}. Moreover, incomplete or incorrect implementations of these solutions make them vulnerable to location discovery attacks. Mobile operating systems also try to prevent undesired location tracking by implementing permissions that all apps must request for accessing location data. These measures, however, are not very effective in preventing location tracking because users are unaware of an app's privacy practices and are often careless about granting such permissions. Also, no protections exist against sensor side-channels (e.g., from Accelerometers and Gyroscopes) even when they are now known to leak location information~\cite{han2012, PowerSpy, nawaz2014, NarainVBN2016, PinMe}.

An alternative protection against location tracking is the generation of synthetic location trajectories~\cite{Bindschaedler2016, Machanavajjhala2008} that are independent of users real locations~\cite{krumm2009, chow2009}. These trajectories guarantee location privacy because it is not possible to derive the user's location from them, however, they risk denial of service if an adversary detects that the trajectories are fake. To be effective against detection, these trajectories must emulate real movements and routes by incorporating real user transitions, movement schedules, traffic information and driving behavior. Synthetic, yet realistic, mobility trajectories are important as they have the potential to eliminate privacy leaks and also enable the understanding of how users' location information is exploited by mobile apps. 

\subheading{Approach} 
The proposed \system~system is designed to address privacy protection weaknesses in Android. To detect leakage from location and sensor data, it implements a \privo~service to monitor and analyze apps patterns for accessing location and sensor APIs. \privo~provides users with real-time notifications and a graphical interface to display how apps access their location information and permissionless sensors (e.g., the time of location access, the accuracy of the location data received, the rate a sensor was sampled, and whether the app was in foreground or background). The service is designed to hide all the underlying complexity from the users and provide them an intuitive interface to help them make more privacy informed decisions about providing synthetic location data to apps using \system~or uninstalling/disabling apps they do not trust. \privo~also implements a permission-protected API that allows security apps installed on the device to get real-time information about which apps access private location and sensors information.

To protect against leakage of location data, MATRIX implements a Synthetic Location Service that gives users the capability of setting their privacy preferences for each installed app. The service dynamically and seamlessly sandboxes apps installed on the device to receive obfuscated or synthetic feeds as specified by the user. The synthetic feeds are generated such that they are difficult to distinguish from real ones by an adversary. To this end, we model user identities and their movements between locations through Finite State Machines (FSM) with probabilistic transitions connecting states. The transitions between states represent routes that are generated from graphs constructed from real road networks. These synthetic routes are made realistic by generating a randomized schedule (path in the FSM) using Linear Programming that satisfies each state's preferences in terms of time spent, and expected arrival in those states. We further incorporate traffic information from historical traffic APIs such as Google Maps Directions API, generate accelerations and speeds using Quadratic Programming based on statistical information from user driving behavior, and also add noise to the synthetic data to emulate real GPS data, in addition to incorporating walk times and idle times. 

\subheading{Contributions} 
Our contributions are as follows:

\begin{itemize}
\item \system~is the first system, to the best of our knowledge, to implement an efficient and extensible auditing system for the Android ecosystem. It audits all location and sensor accesses by all apps on the device to detect leakages, generates real-time notifications and graphs for visualizing these accesses in an easy and intuitive manner. 

\item \system~gives users the capability to change their privacy preferences and provide obfuscated or synthetic trajectories to installed apps. It is the first system, in our knowledge, to generate realistic synthetic identities and trajectories to protect users' privacy. We show that generating such trajectories is feasible by incorporating traffic information. The trajectories are randomized yet satisfy realistic schedule constraints using a randomized linear program, and match statistical characteristics of user driving behavior using a quadratic program.

\item MATRIX is an extensible system integrated within Android without modifications to the operating system, nor requires rooting. It will be extended to incorporate other sensitive APIs, e.g., Wi-Fi, Telephony, Camera and Microphones to provide users a holistic view of accesses to their private information. It can also be used by security apps and researchers to identify which apps misuse/leak private location and sensors information, by analyzing an app's accesses and injecting synthetic honey-data to observe if it is used in contexts not authorized by users. 

\item We extensively evaluated \system~to validate system performance and reliability, and realism of synthetic trajectories. Testing 1000 popular Android apps, we report negligible impact in performance and reliability. For 10 popular location-driven apps, we report that MATRIX is undetected while at least one app could detect non-MATRIX mobility patterns. Our user study involving 100 users indicates that the synthetic trajectories are difficult to differentiate from real traces visually, with more users confusing synthetic trajectories to be real. Our machine learning evaluation indicates that most well-known algorithms fail to differentiate between real and synthetic trajectories with an average accuracy of 50\% (comparable to an algorithm that uses a coin-flip), with just one algorithm achieving an accuracy of 63\% in guessing if a trajectory is synthetic.
\end{itemize}

%The paper is structured as follows. In~\Cref{sec_mitigation_background}, we discuss the current Android location privacy protection mechanisms. In~\Cref{sec_mitigation_approach}, we describe the high-level design of \system~and how it addresses weaknesses in the protection mechanisms. \Cref{sec_mitigation_solution} and \Cref{sec_mitigation_synthetic} provide a detailed description of \system's architecture and the technique for generating realistic privacy-preserving synthetic user identities and mobility trajectories. In~\Cref{sec_mitigation_evaluation}, we present the results of our evaluations. In~\Cref{sec_mitigation_relatedwork}, we describe some of the related works. We conclude in \Cref{sec_mitigation_conclusion}.

\section{Location Privacy in Android}
\label{sec_mitigation_background}

This section provides a background on Android location and sensor APIs, the current Android privacy protection schemes and their weaknesses.

\subsection{Android Location \& Sensor APIs}
\label{sec_mitigation_apis}

The \system~system audits all location and sensor accesses and updates the location information reported to an app in some contexts. There are a standard set of Android APIs that provide this information.

Location information can be accessed using four different APIs. The \code{LocationManager} is the default API available in all versions of the Android SDK. The \code{FusedLocationProviderClient}, \code{FusedLocationProviderApi} (deprecated) and \code{LocationClient} (deprecated) are provided by Google Play services as recommended closed source alternatives that consume less battery for higher accuracy data. All these APIs contain \code{request*} and \code{remove*} calls (e.g., \code{requestLocationUpdates} in \code{LocationManager}) that enable apps to register and unregister for continuous location updates. Once registered, location information is sent asynchronously to the listeners based on the criteria set by the app (e.g., quality, rate, latency). These managers also contain additional methods such as  \code{getLastKnownLocation} in \code{LocationManager} that can return a location update immediately.

Sensor information (e.g., Accelerometers and Gyroscopes) can be accessed using the \code{SensorManager} API. It is important to note that access to these sensors does not require permissions in any versions of Android. Also, these sensors can  be accessed by apps in the background without any notification or visual cues to the user.

\subsection{Weaknesses in Privacy Protections}
\label{sec_weaknesses}

Android implements some location privacy protection schemes to give users the capability to control how and whether certain apps can access their location information. These schemes are not sufficient for protecting a user's privacy. Some of the weaknesses are discussed below. Note that these weaknesses are labeled (\textbf{W\#}) for ease of referring to them in the next section.

\subheading{Weak Permissions Model (W1)}
Android specifies two permissions for limiting access to the user's location information: \code{ACCESS\_FINE\_LOCATION} and \code{ACCESS\_COARSE\_LOCATION}. The former allows apps to access high accuracy location information, while the latter provides obfuscated information to hide the user's real location. The permissions model is a good step in notifying users of location access, however, this protection is limited as users have an option to always allow access. This means that the user will not be notified about location access again even if the app's context has changed, i.e., location is accessed from another activity or from a service, or a previously benign app is updated with a privacy intrusive version.

\subheading{Non-existent Auditing Capabilities (W2)}
Android does not provide a framework to audit how apps access a user's private information. Also, App stores (e.g., Google Play Store) do not provide enough information about the privacy practices of an app. Without any privacy-related knowledge, users are more than likely to install and use an app if they require the services provided by that app.

\subheading{Weak Location Activity Notification (W3)} 
The Android operating system displays a notification icon on the notification bar of the device, whenever any app requests continuous location updates. An adversary can easily bypass this protection by using an alternative method for location access. One example is the \code{getLastKnownLocation} call in \code{LocationManager} which can be invoked numerous times for receiving continuous location updates. Another example is exploiting the permissionless sensors like Accelerometers, Gyroscopes and Barometers to infer user locations. Both these methods do not display any notifications of access. Furthermore, the notification simply indicates that some app has access to location and no further information is given to the user to make privacy-aware decisions.

\subheading{Restricted Privacy Preferences (W4)}
Android does not provide the capability for users to define their privacy preferences for apps installed on their device. Users can deny location access to certain apps by disallowing location permissions, however, certain apps may then deny service to the users. There are situations in which users may not wish to disclose their locations, in particular at some moments in time, and still require the services of the app. One example of this is when the app is turned-off or in the background.

\subheading{Weak Location Granularity Settings (W5)}
Android implements a location obfuscation scheme that hides a user's real location from apps requiring just the \code{ACCESS\_COARSE\_LOCATION} permission. This obfuscated location still leaks information about the user's location. There is currently no mechanism for users to completely hide their location by providing synthetic information to untrusted apps.

\section{High-Level Approach}
\label{sec_mitigation_approach}

\system~is an extensible system designed to address several location privacy protection weaknesses in Android. It uses the Android design paradigms for easy integration into the Android ecosystem with minimal changes. \system~hides all implementation complexity from the end-users to make the system easy to use and intuitive. The system comprises of three modules: an App-activity \privo~Service, a Synthetic Location Service, and a Synthetic Location Provider. 

The App-activity \privo~Service monitors and analyzes apps patterns of location and sensor API accesses. It is designed for end-users, researchers and security apps desiring to assess the privacy posture of installed apps on the device. To the best of our knowledge, this service is currently the only one to provide an efficient auditing capability in Android. End-users can view all location and sensor access information as intuitive graphs.  Other apps can get real-time audit events via a permission protected secure API \textbf{(W2)}. The service also displays real-time visual notifications of location and sensor access activity to users. The notification bar is updated whenever any app accesses these sensors and displays information about which apps are actively accessing what sensors on the device \textbf{(W3)}. The architecture of \privo~is described in \Cref{sec_mitigation_privoscope}.

The Synthetic Location Service provides an interface to the user to set their location privacy preferences for all installed apps. This service implements three settings for location granularity: Block level, City level and Synthetic Locations \textbf{(W4)}. The architecture of this service is described in \Cref{sec_mitigation_synservice}. The last module, the Synthetic Location Provider provides the Synthetic Location Service obfuscated/synthetic locations whenever the service requests for it. To the best of our knowledge, this provider is currently the only one to generate realistic privacy-preserving synthetic identities and mobility trajectories for protecting users' privacy \textbf{(W5)}. The techniques for modeling synthetic identities and movements are described in \Cref{sec_mitigation_synthetic}.

The Synthetic Location Service relies on the default Android permission manager for managing location permissions, however, it restricts location access to background apps by default. Instead of completely denying location information, it detects if the requesting app is in the background and provides it the last location fix that the app received in foreground to prevent it from tracking users \textbf{(W1)}.
  
\begin{figure}[t]
\begin{center}  
\includegraphics[width=\linewidth]{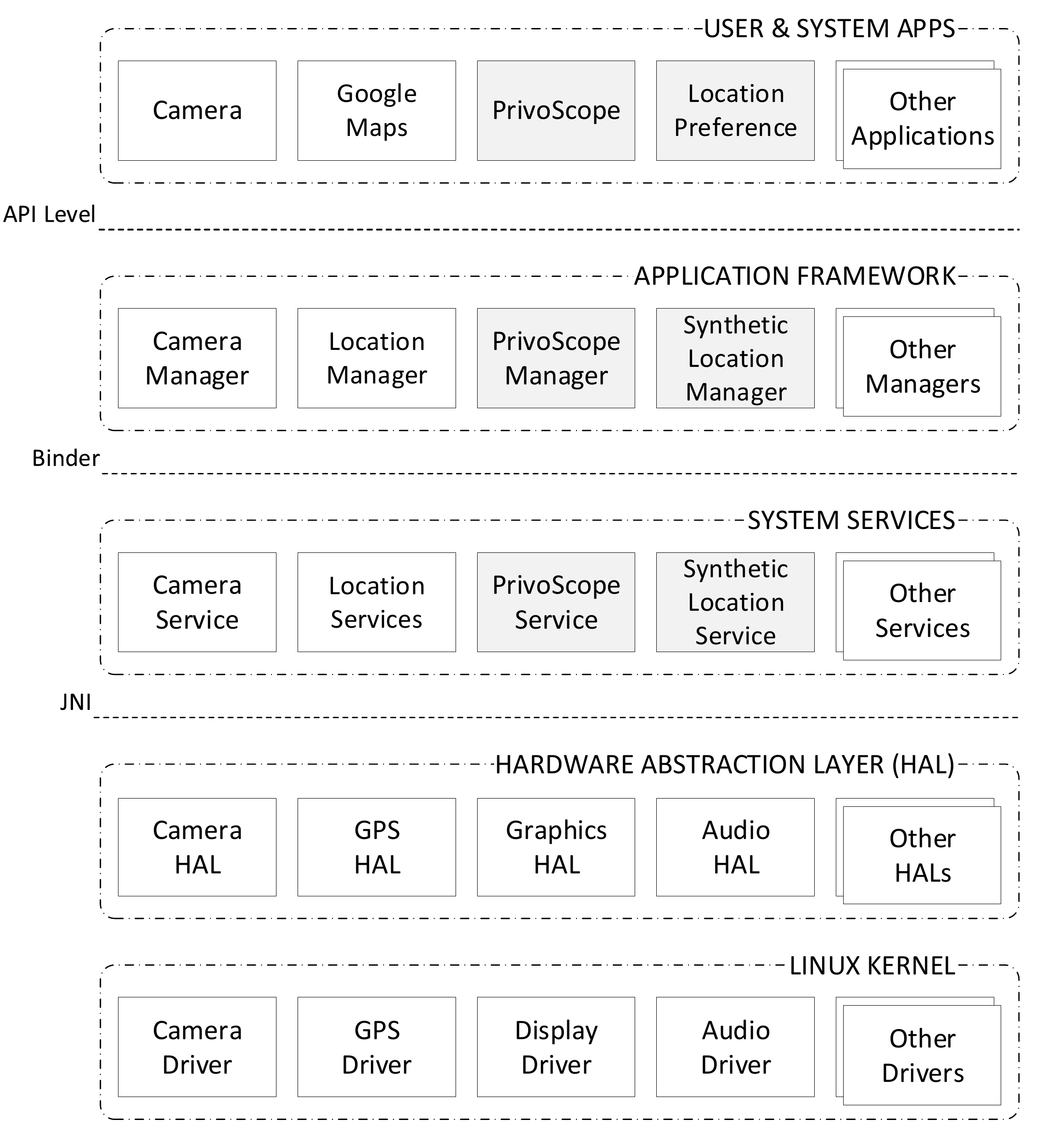}
\caption{\system~integration into the Android ecosystem.\label{fig_matrixstack}}
\end{center}  
\end{figure} 

\Cref{fig_matrixstack} shows how \system~integrates into the Android ecosystem. The PrivoScope Service and Synthetic Location Service are implemented as system services that start at device boot and are registered in the system server registry. These services implement all the protections that ensure that only authorized apps can use their functions. Apps installed on the device interact with these services using APIs provided by the PrivoScope Manager and Synthetic Location Manager. These managers are loaded into each app's process and communicate with the corresponding services. At the user level, \system~implements a PrivoScope GUI that provides a graphical interface to the users to analyze the app's privacy practices, and a Location Preference GUI that enables users to set their location privacy preferences. These also use the PrivoScope and Synthetic Location Managers to communicate with the corresponding system services.

\section{\system~Architecture}
\label{sec_mitigation_solution}

This section describes the architecture of the \privo~and Synthetic Location services implemented for the \system~framework.

\subsection{API Call Interception} 

Previous mitigation systems (excluding Boxify~\cite{boxify}) were implemented by either modifying the Android source code, using rooted devices, or using third party frameworks such as the Xposed Framework \cite{xposed}. The Xposed framework adds an extended \code{app\_process} executable in the \code{/system/bin} folder of the device on installation. This extended \code{app\_process} adds an additional \code{jar} file to the classpath and calls methods even before the \code{main} method of \code{Zygote} is called. This enables apps to intercept method calls that are otherwise inaccessible from an app's process.

\system~uses the Xposed framework to intercept location and sensor API calls. One example usage in our context is intercepting the \code{requestLocationUpdates} method of \code{LocationManager} to generate an event every time an app requests location updates. This event contains all the relevant information about the request and sent to \privo~for logging and notification. Using the framework is both necessary and advantageous due to the following reasons: (1) Xposed has the capability to intercept external APIs like Google Play Services which is currently not possible by modifying the Android source or by rooting, (2) the framework is supported and has a consistent API for different versions of Android ensuring portability and ease of development, and (3) the framework does not require a rooted device to function properly. We developed a simple tool that automates the installation of Xposed and \system~through a custom recovery (e.g., TWRP \cite{twrp}) without rooting the device. The Xposed framework and TWRP recovery are both open-source and consistently analyzed and updated by a large community of Android users, making them quite reliable.

\subsection{The App-activity PrivoScope Service}
\label{sec_mitigation_privoscope}

\begin{figure}[t]
    \centering
    \begin{subfigure}[b]{0.24\textwidth}
        \centering
        \includegraphics[width=\linewidth]{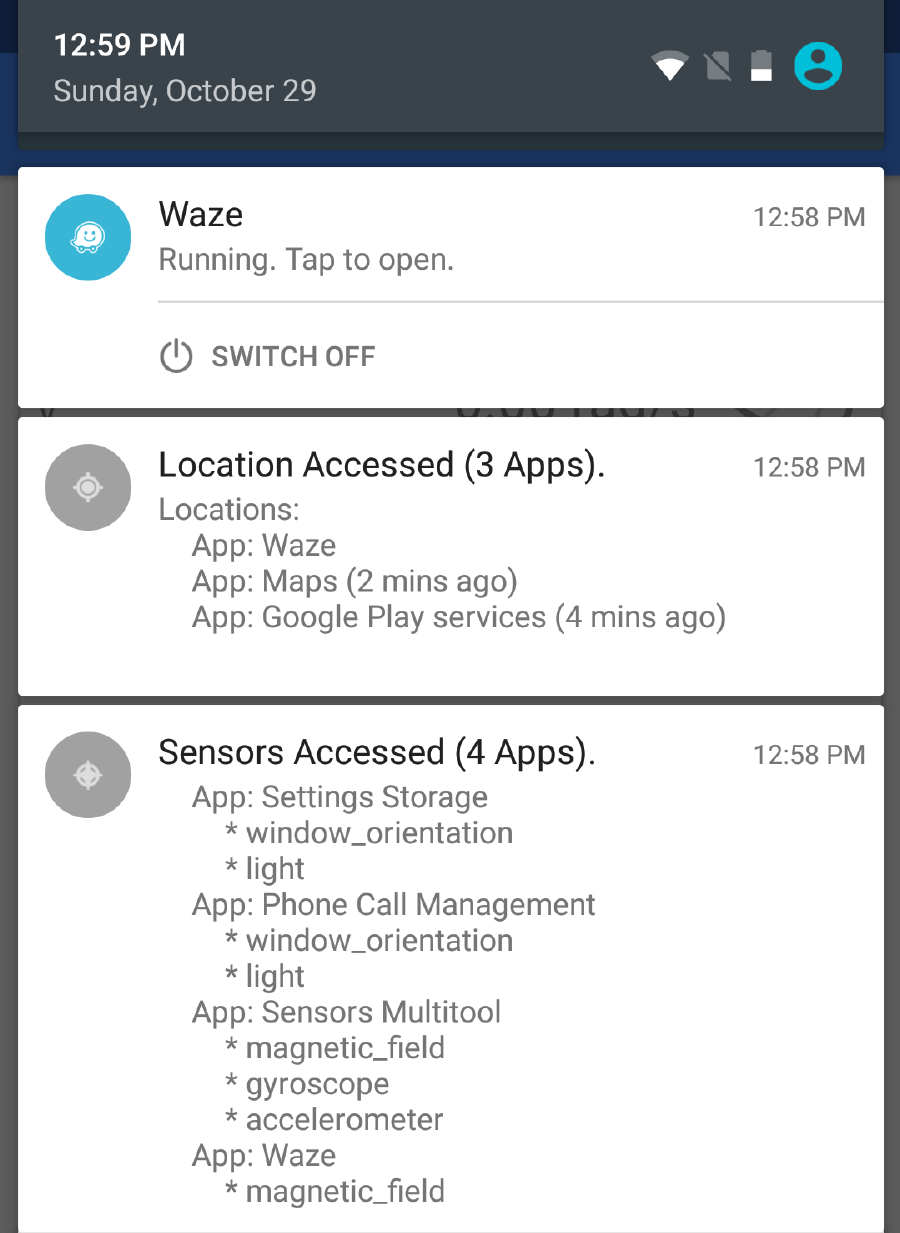}
        \caption{\label{fig_privo_1}PrivoScope Notification Bar}    
    \end{subfigure}
    \begin{subfigure}[b]{0.24\textwidth}  
        \centering 
        \includegraphics[width=\linewidth]{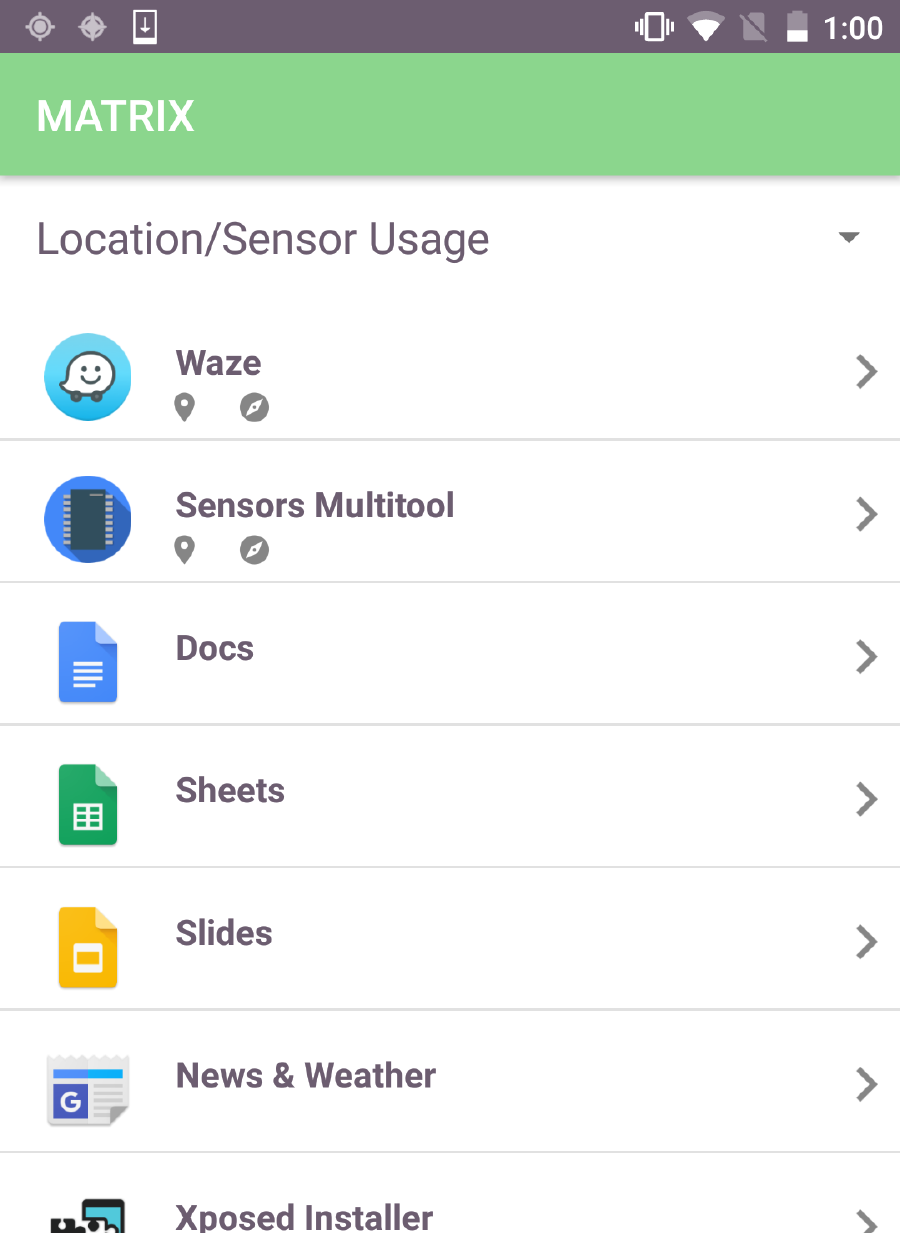}
        \caption{\label{fig_privo_2}App Selection Activity}    
    \end{subfigure}
    \begin{subfigure}[b]{0.24\textwidth}   
        \centering 
        \includegraphics[width=\linewidth]{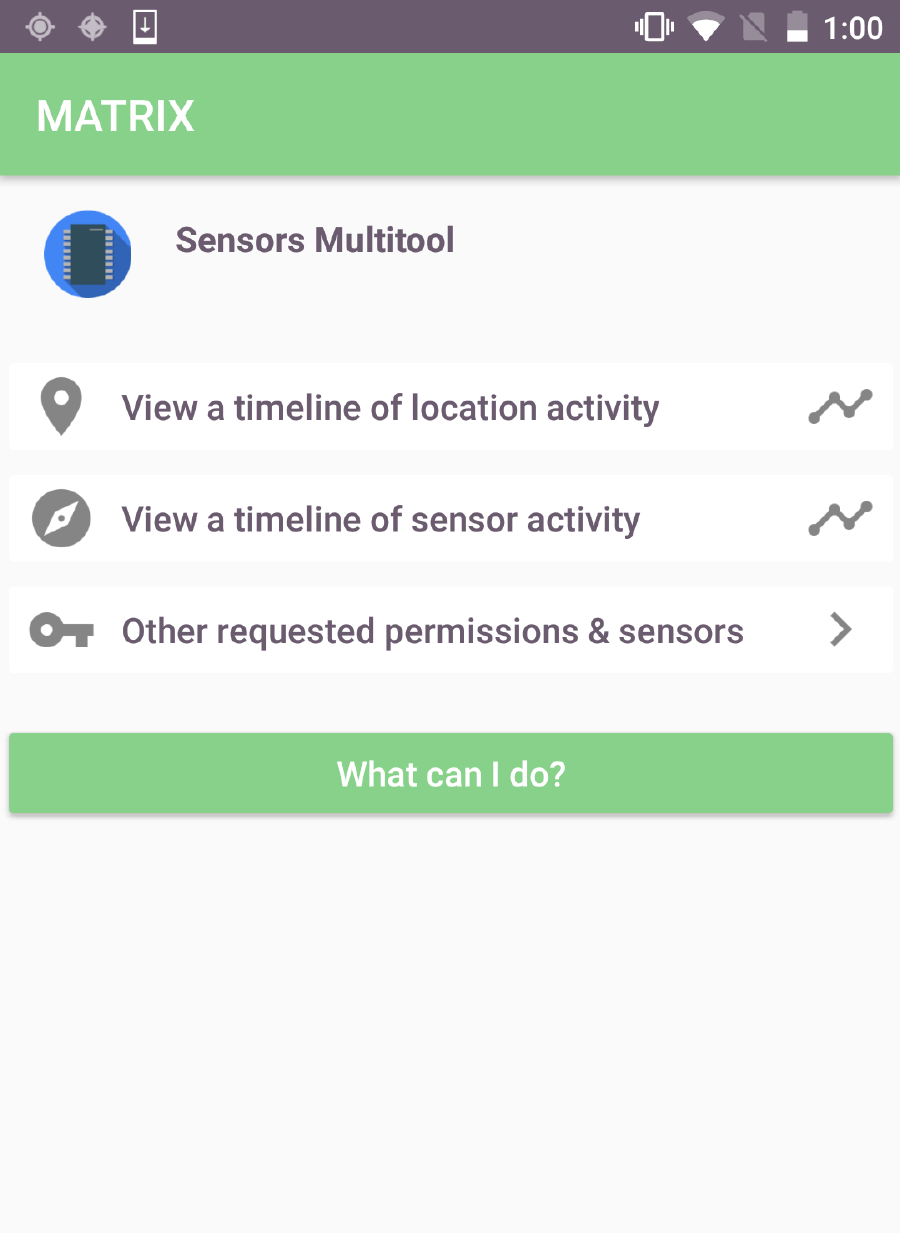}
        \caption{\label{fig_privo_3}App Detail Activity}    
    \end{subfigure}
    \begin{subfigure}[b]{0.24\textwidth}   
        \centering 
        \includegraphics[width=\linewidth]{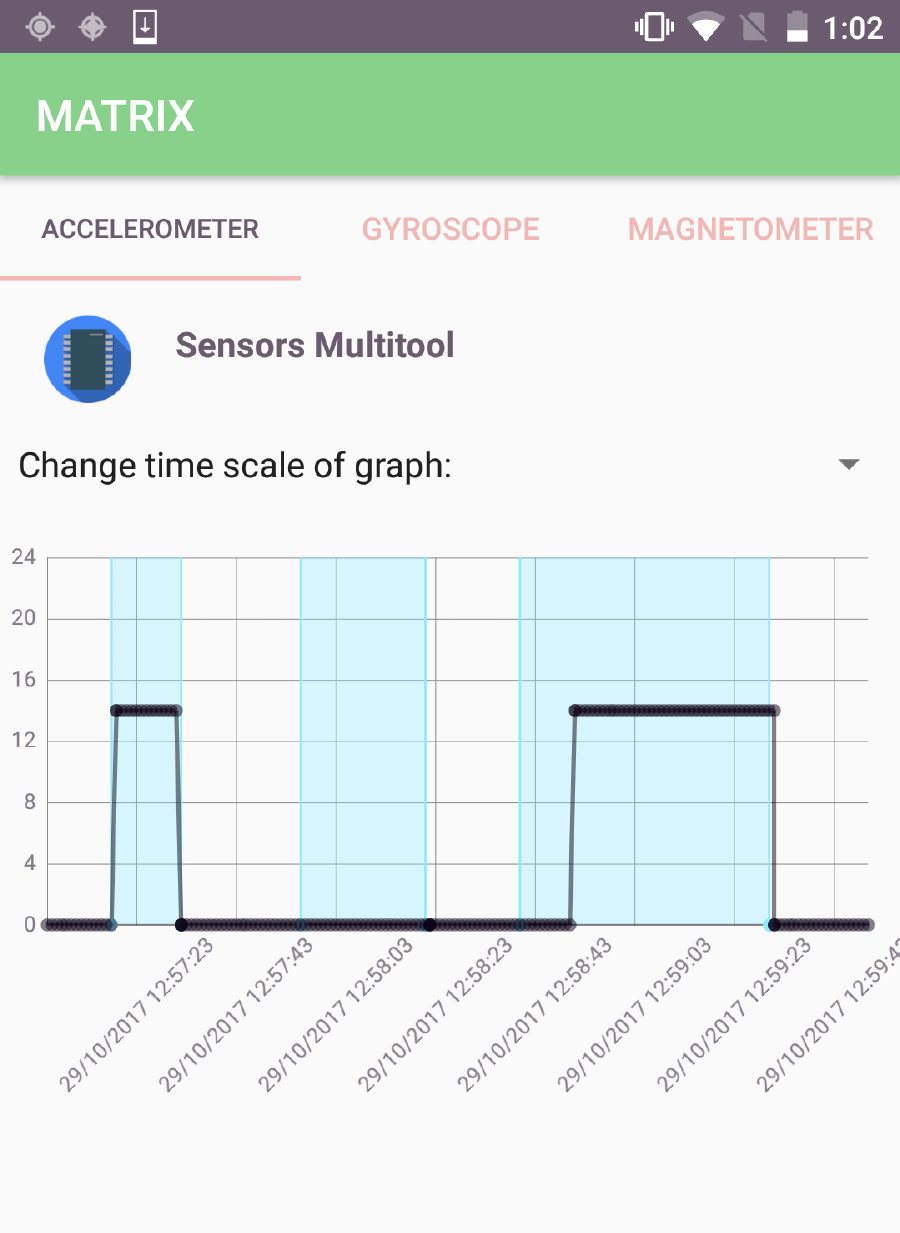}
        \caption{\label{fig_privo_4}Accelerometer Timeline}    
    \end{subfigure}
    \caption{Example screenshots of the PrivoScope GUI.} 
    \label{fig_privo}
\end{figure}

At a high level, the \privo~service uses the Xposed framework to intercept all location and sensor APIs, generates events containing the audit details, adds the events to a database and displays real-time usage notifications to the end-user. The service also exposes a permission protected API that other security apps can register to get real-time and archived audit events. It also implements a GUI interface for the users to analyze app behavior on their device. The motivation is to help users make privacy aware decisions regarding installed apps. \Cref{fig_privo} shows example screenshots of the PrivoScope GUI, where \Cref{fig_privo_1} shows the PrivoScope real-time location and sensor usage notification, \Cref{fig_privo_2} shows a list of installed apps sorted by most recent access of location and sensor APIs, \Cref{fig_privo_3} shows links to an app's permissions and access details, and \Cref{fig_privo_4}  shows a timeline of Accelerometer access by an app at different times. This timeline can be set to display accesses in the past month, week, day, or a custom number of hours. Note that an app's life-cycle is color coded to help users differentiate between foreground and background accesses. Here, blue indicates that the app was in the foreground while gray would indicate background access. The evaluation and performance analysis of PrivoScope is reported in \Cref{sec_performance}.

\begin{figure}[t]
    \centering
    \includegraphics[width=.9\linewidth]{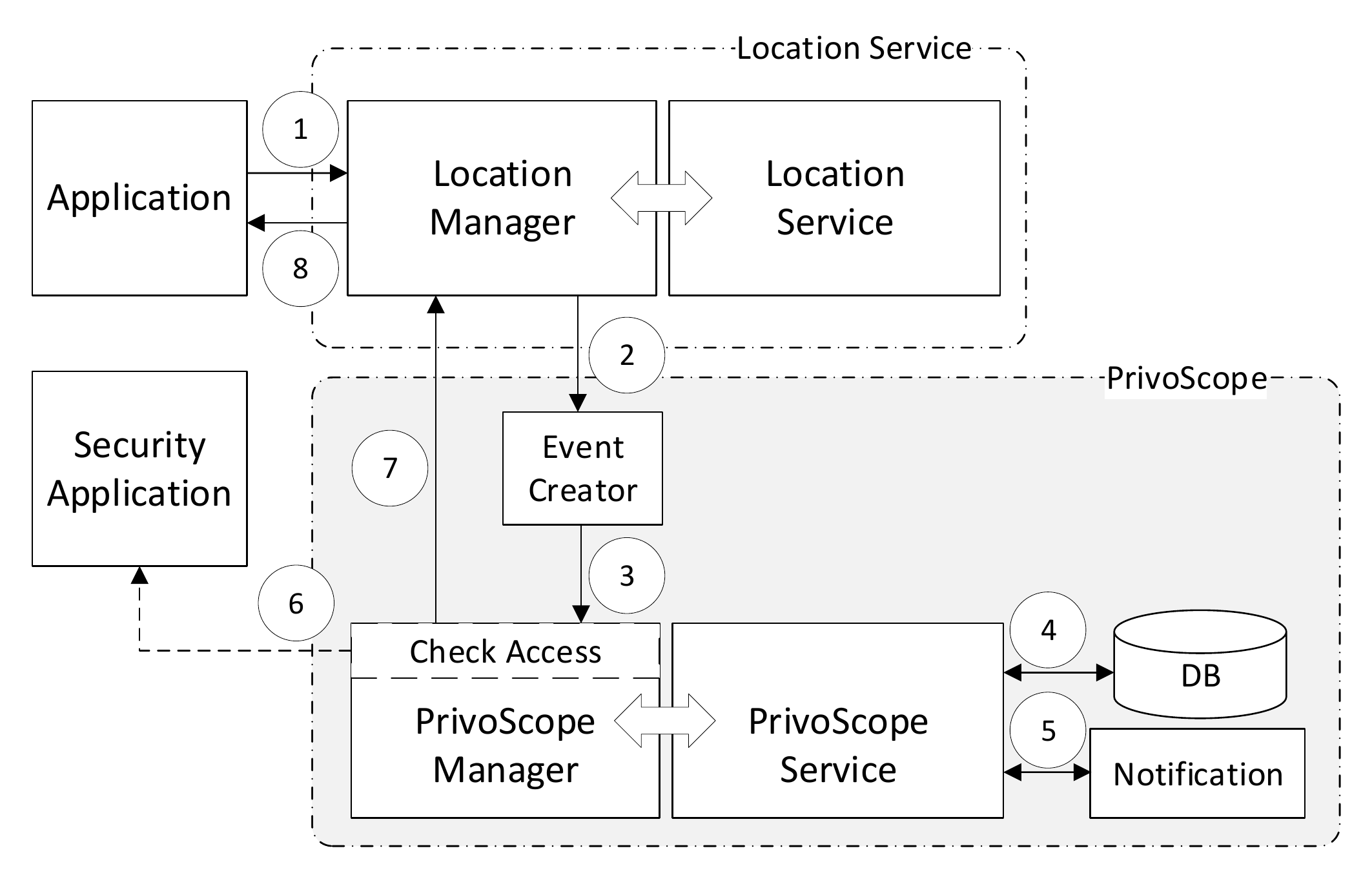}
    \caption{Architecture diagram of the PrivoScope service.}
    \label{fig_arch1}
\end{figure}

The architecture of the PrivoScope service for a \code{requestLocationUpdate} method call from \code{LocationManager} is shown in \Cref{fig_arch1}. Note that we abstract away from the low level implementation details and Android's internal complexities in this paper for simplification. Also note that the architecture is generic across all location managers and the sensor manager and we use \code{LocationManager} here just for illustration purposes. Like all other Android services, \privo~implements a manager called \code{PrivoScopeManager} that exposes public APIs to other apps and a system service called \code{PrivoScopeService} that performs all the security sensitive operations and checks if apps have appropriate access rights for their services. 

The control flows like this: An app requests continuous location updates using the \code{requestLocationUpdate} method call from \code{LocationManager}. The manager and the privileged \code{LocationManagerService} validate the app's access by checking its requested permissions \textcircled{1}. Once access is validated, the API call interception service generates an event containing all relevant information to be logged for auditing. All private user information contained by the request are ignored. For example, this specific event would contain \textit{the system time, the app package name, the activity invoking the request, whether the app is background or foreground, the requested location provider, and the requested accuracy and sampling rate} \textcircled{2}. This event is then sent to the \code{PrivoScopeManager} for logging using an \code{addAuditEvent} method call exposed by the manager \textcircled{3}. The \code{PrivoScopeManager} forwards this event to the \code{PrivoScopeService} which validates whether the package name in the event is the same as the package name of the app making the request. This ensures security as only apps generating an event can add the event. The event is discarded if the package names do not match and a \code{SecurityException} is thrown. In case of a successful match, the event is added to the service's database \textcircled{4}. The \code{PrivoScopeService} also sends this event to a Notification service that keeps track of all active apps accessing location and sensor APIs and updates the notification bar with this new event information \textcircled{5}. The \code{PrivoScopeManager} exposes a \code{requestAuditEvents} method call that other apps on the device can register for receiving real-time audit events. This call is protected using a custom permission called \code{GET\_AUDIT\_EVENTS} and apps must request this permission for access. The \code{PrivoScopeManager} sends the event to all registered apps that receive this event asynchronously using a \code{AuditEventListener} callback interface \textcircled{6}. Based on whether this event was successfully added to the database or not, the \code{addAuditEvent} method call returns a boolean value to the \code{LocationManager} \textcircled{7}. Note that steps \textcircled{3} to \textcircled{7} execute in a new thread to ensure that the app functionality and the performance is not impacted by PrivoScope. After step \textcircled{3}, the \code{requestLocationUpdate} method call simply terminates as its return type is a \code{void}. The other method calls and managers return the expected values and their functionality is not updated by PrivoScope \textcircled{8}. 

\subsection{The Synthetic Location Service}
\label{sec_mitigation_synservice}

\begin{figure}[t]
    \centering
    \includegraphics[width=.9\linewidth]{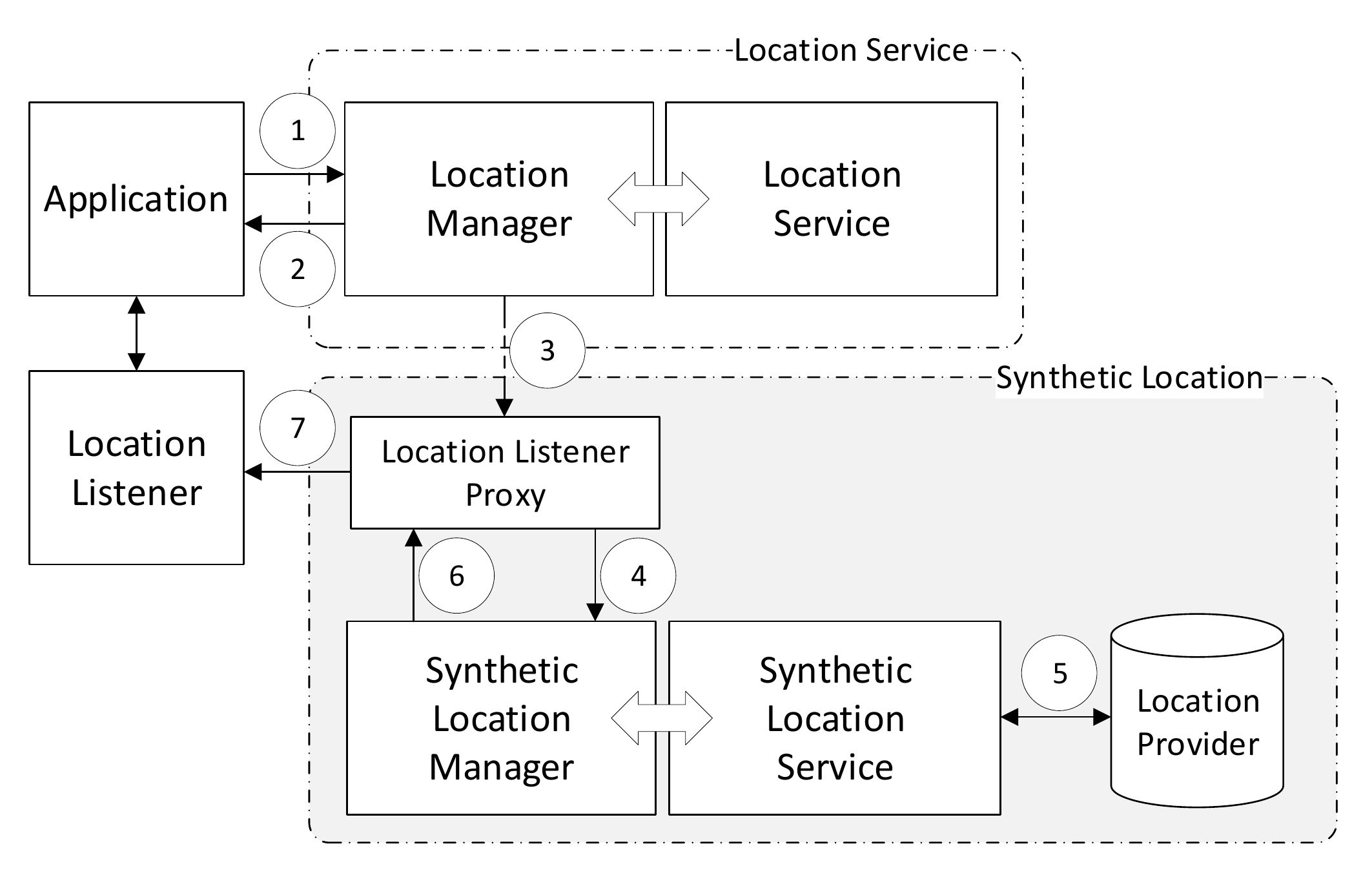}
    \caption{Architecture diagram of the Synthetic Location service.}
    \label{fig_arch2}
\end{figure}

The architecture of the Synthetic Location service is shown in \Cref{fig_arch2}, again in the context of receiving location updates from the \code{LocationManager} API. Like \privo, this architecture is generic across all the location managers. The Synthetic Location service implements a manager called \code{SyntheticLocationManager} that exposes public APIs to other apps and a system service called \code{SyntheticLocationService} that manages and protects the database storing the user location preferences, and connects with the \code{LocationProvider} to request obfuscated/synthetic locations. 

The control flows like this: When an app requests continuous location updates (with the correct permissions) using the \code{requestLocationUpdates} call from \code{LocationManager}, the first steps that occur are the listener registration (cf. \Cref{sec_mitigation_apis}) and addition of the audit event to the PrivoScope service's database (cf. \Cref{sec_mitigation_privoscope}). \textcircled{1}, \textcircled{2}. After registration is completed, all the location fixes generated by the \code{LocationManagerService} are typically sent asynchronously to the app's \code{LocationListener}, \code{PendingIntent} or \code{LocationCallback} implementation.  In \system, these location fixes are intercepted by a \code{LocationListenerProxy} that proxies it to the app's listener. The proxy works by hooking the \code{Location} object that is used by all the managers to send location fixes to the app's listener. This enables it to modify the location object before the app loads the information using the get* method calls (e.g., \code{getLatitude() and getLongitude()})  \textcircled{3}. The \code{LocationListenerProxy} requests the \code{SyntheticLocationManager} to provide an updated location for the app, based on the app's location preference set by the user. The manager forwards this request to the \code{SyntheticLocationService} that maintains and protects the database storing the user location preference for each app \textcircled{4}. The \code{SyntheticLocationService} looks up the user's location preferences in the database, and communicates with the \code{LocationProvider} to request an obfuscated/synthetic location if the user has chosen to receive such location information for the app. The default preference set for an app requesting fine location is block level obfuscated data ($500m$) \textcircled{5}. An updated location object is returned to the \code{SyntheticLocationService} which forwards it to the \code{SyntheticLocationManager}. The \code{SyntheticLocationManager} sends this location to the \code{LocationListenerProxy} that updates it before the app accesses the location \textcircled{6}, \textcircled{7}.

\begin{figure}[t]
    \centering
    \begin{subfigure}[b]{0.24\textwidth}
        \centering
        \includegraphics[width=\linewidth]{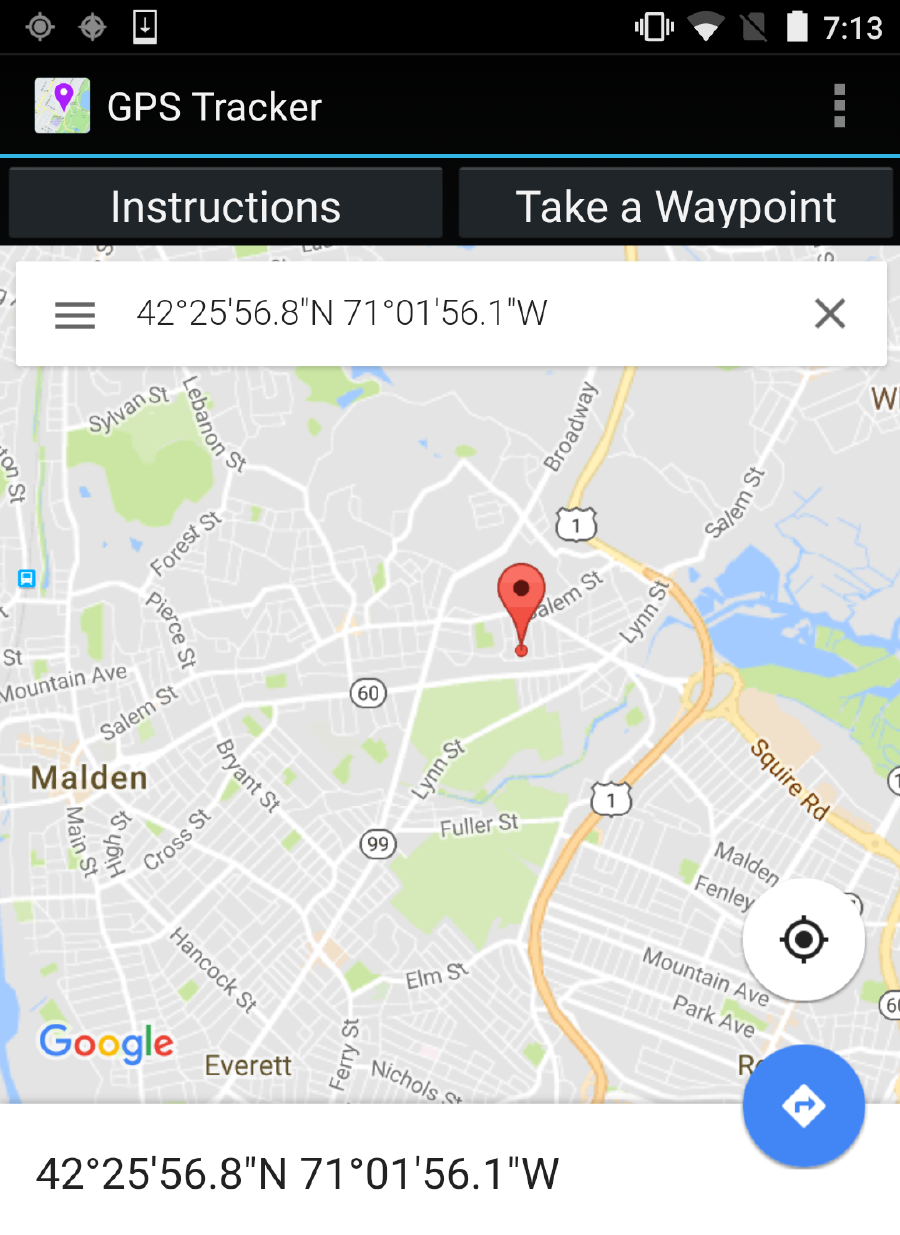}
        \caption{\label{fig_syn_1}Real Location}    
    \end{subfigure}
    \begin{subfigure}[b]{0.24\textwidth}  
        \centering 
        \includegraphics[width=\linewidth]{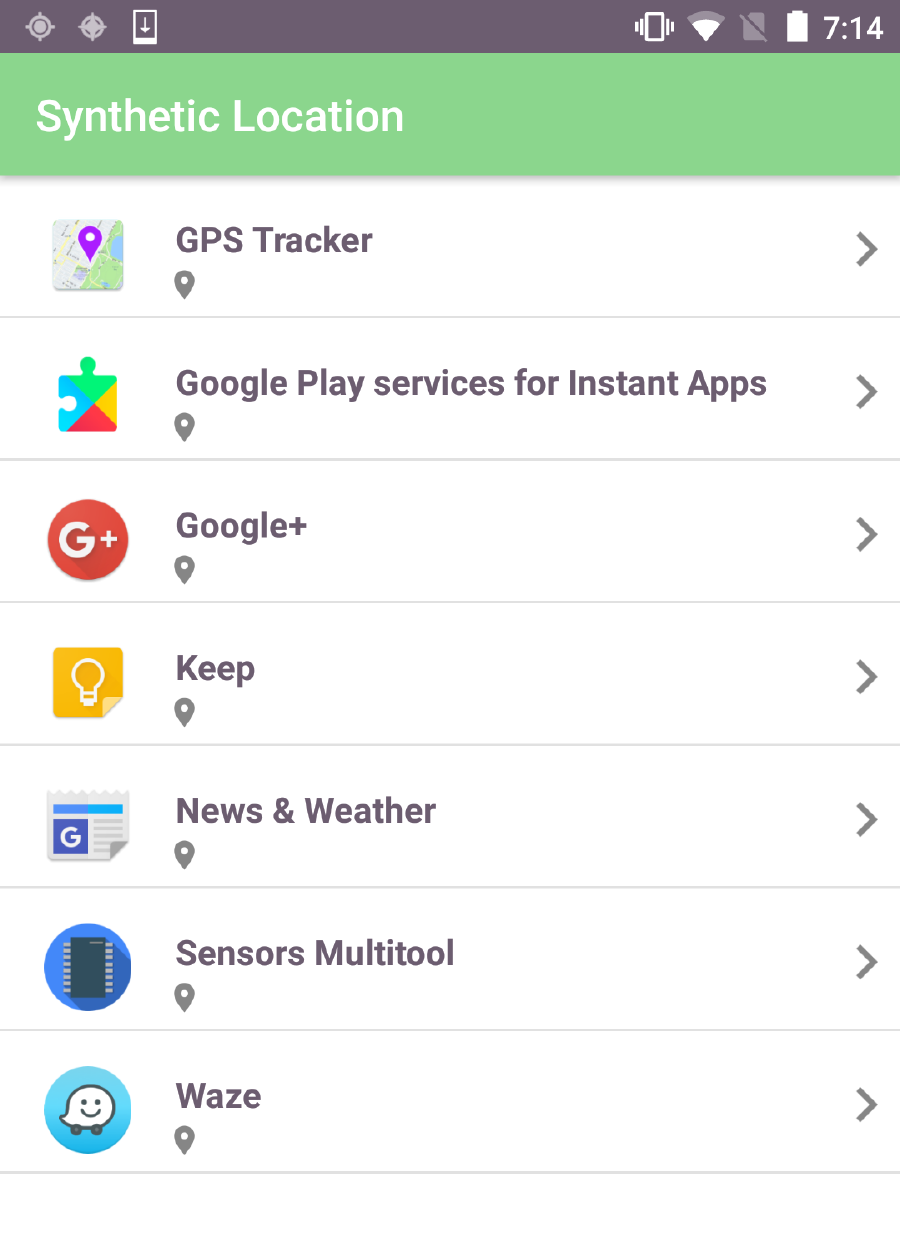}
        \caption{\label{fig_syn_2}App Selection Activity}    
    \end{subfigure}
    \begin{subfigure}[b]{0.24\textwidth}   
        \centering 
        \includegraphics[width=\linewidth]{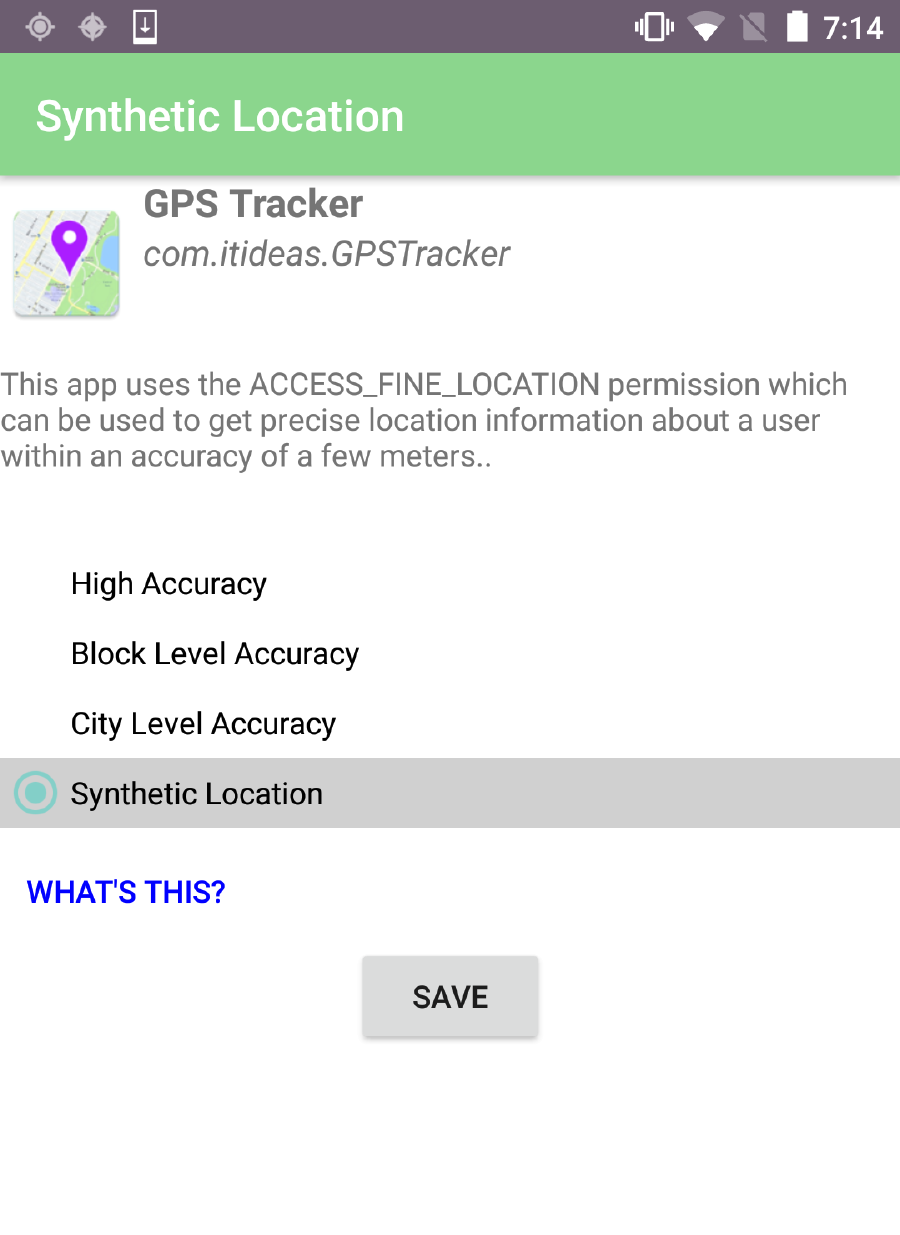}
        \caption{\label{fig_syn_3}Location Preference}    
    \end{subfigure}
    \begin{subfigure}[b]{0.24\textwidth}   
        \centering 
        \includegraphics[width=\linewidth]{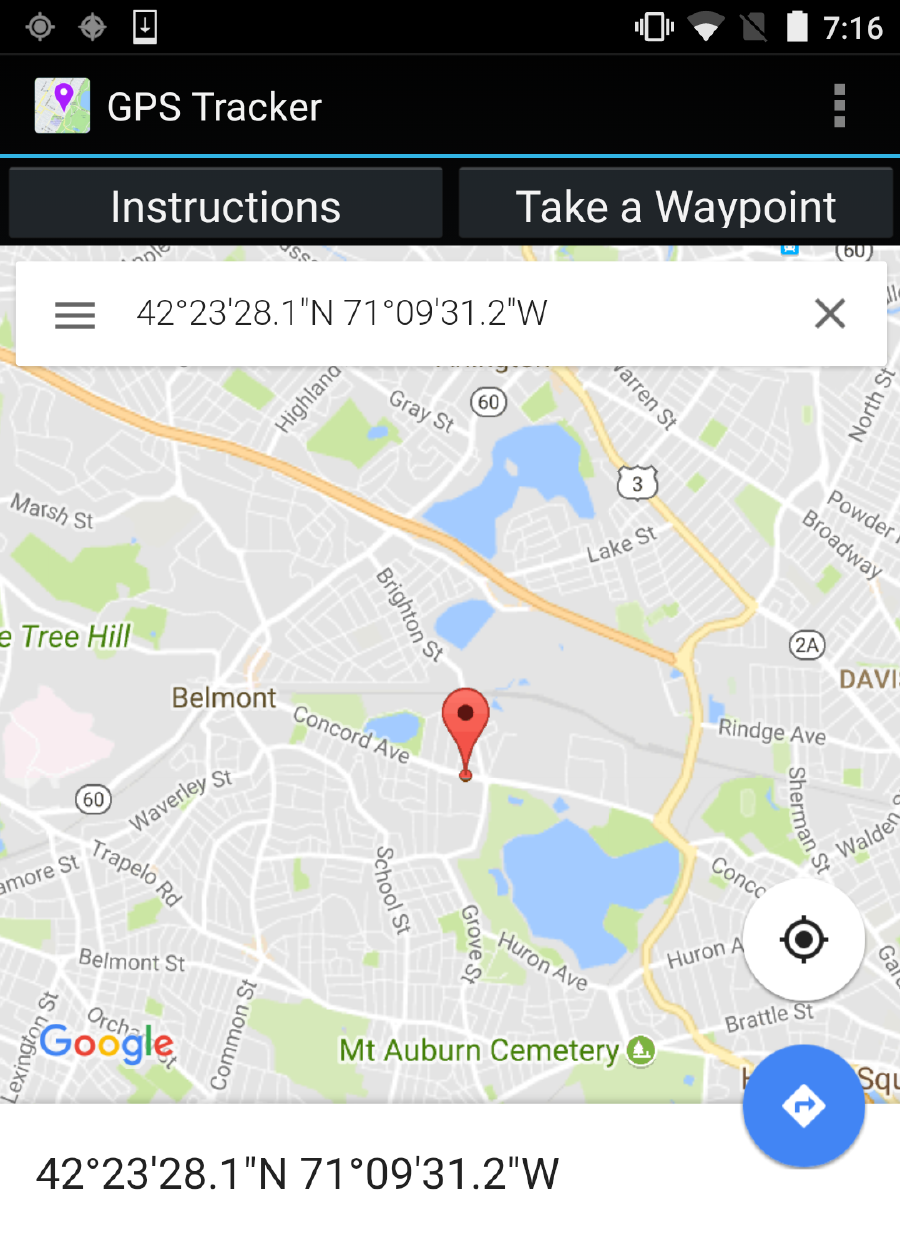}
        \caption{\label{fig_syn_4}Synthetic Location}    
    \end{subfigure}
    \caption{Example screenshots of the Synthetic Location GUI.} 
    \label{fig_syn_ui}
\end{figure}

The Synthetic Location service currently provides four settings for per-app location privacy: High Accuracy, Block Level Accuracy, City Level Accuracy, and Synthetic Locations. Note that the High Accuracy and Block Level Accuracy options are only available for apps requesting fine location using the \code{ACCESS\_FINE\_LOCATION} permission. This is because apps that use \code{ACCESS\_COARSE\_LOCATION} permissions already receive coarser location data than that provided by the two options. The high accuracy option set for an app tells the service to not obfuscate or synthesize locations for this app. For block level and city level accuracy, we extended the default Android \code{LocationFudger} implementation to support different grid resolutions. The implementation is in \code{com.android.server.location.LocationFudger} under the Android source tree~\cite{aosp}. We analyzed this code to find that the real location information is obfuscated in two steps. First, a random offset is applied to the location to mitigate against accurate detection of grid transitions when a user crosses a grid boundary. This offset is changed slowly over time (e.g., once every hour) to mitigate against location inference attacks. Second, the primary means of obfuscation is to snap the offset data (already mitigated against grid transitions) to a grid. This grid radius chosen by most recent versions of Android is $2000m$. We found this technique to be effective against location inference attacks. The current grid radius settings for block level and city level accuracy are $500m$ and $5000m$, respectively. 

\Cref{fig_syn_ui} shows screenshots, illustrating the Synthetic Location service for a GPS tracking app. Note that this app is used for demonstrating how the service works because it displays the user location on the screen, and it is not a malicious app. \Cref{fig_syn_1} shows the test app displaying the user's real location, \Cref{fig_syn_2} shows the list of installed apps that request location permissions, \Cref{fig_syn_3} shows the location privacy preference for the test app being changed to synthetic, and \Cref{fig_syn_4} shows the test app now displaying a synthetic location in another city.  This synthetic location is provided based on the time of the day and a realistic GPS trajectory generated for the user for that specific day.

\section{Generating Synthetic Identities}
\label{sec_mitigation_synthetic}

This section provides a detailed description of our technique for generating unique and realistic privacy-preserving synthetic identities and mobility trajectories for each user using the \system~system. 

\subsection{Modeling User States}
\label{sec_states}

A user's synthetic mobility patterns are defined as an automated probabilistic state machine with a finite set of $S$ states $Q = \{Q_0, \ldots, Q_{S-1}\}$. The states, in this context, represent a set of tuples \{(Loc($Q_i$), $t_{min, i}$, $a_{min, i}$, $a_{max, i}$)\}, where Loc($Q_i$) is the geographic coordinates of state $Q_i$, $t_{min, i}$ is the minimum time spent in the state, and $a_{min, i}$, $a_{max, i}$ are the lower and upper time bounds for arrival at the state. The geographic coordinates of the states are obtained from OpenStreetMap by parsing the `building' and `amenity' tags \cite{osmbuilding, osmamenity} of all ways and nodes for the given area. For instance, a `Home' state can be chosen as a way or node in OpenStreetMap whose building type is one of the following: `apartments', `house', `residential', or `bungalow'. Similarly, a `Work' state can be chosen from the `commercial' or `industrial' tags. The other attributes are used for scheduling the user's activity for each day and set based on typical times that these activities occur. Note that the attributes are set to default values when they are unimportant for a state, i.e., $t_{min,i}=0$, $a_{min,i}=00$:$00$:$00$, and $a_{max,i}=23$:$59$:$59$. In the simplest form, a state machine may contain just two synthetic states $Q = \{Q_0, Q_1\}$, where $Q_0$ = `Home' and $Q_1$ = `Work'. We label these as \emph{significant} states as the user spends most of their time in one of these states. The geographic coordinates Loc($Q_0$) and Loc($Q_1$) are randomly chosen from the list of all locations with the relevant tags. Assuming no `Work from Home' scenarios, the probabilities $P(Q_0)$ and $P(Q_1)$ of occurrence of these states is taken to be $1$.

The state machine is made more realistic by adding synthetic states like $Q_2$ = `School', $Q_3$ = `Gas Station', $Q_4$ = `Lunch' and $Q_5$ = `Dinner'. We label these as \emph{transitional} states because a user will temporarily visit these states when transitioning between \emph{significant} states (i.e., $Q_0$ and $Q_1$). For any transitional state $Q_i$, the geographic coordinates Loc($Q_i$) is selected from a set of locations Loc = $\{Loc_1, \ldots, Loc_N\}$ with the relevant tags, such that its distance is shortest from the significant states, i.e., $\text{Loc}(Q_i) = \arg\min_{L \in \text{Loc}} d(L, \text{Loc}(Q_0)) + d(L, \text{Loc}(Q_1))$. Note that, unlike \emph{significant} states, visits to \emph{transitional} states are occasional based on some specific frequency of occurrence. This frequency, denoted by $f_i$, is derived from a uniform distribution $\mathcal{U}(l, u)$ with $l$ and $u$ as the bounds for the frequency of visits to that state (e.g., once a week to once a month). In case of `Gas Station' specifically, the system chooses a random mileage $m$ and gas capacity $c$, and calculates the frequency as the number of days a user can travel between the \emph{significant} states before the gas level goes below $1/4^{th}$ of capacity, i.e., $f_3 = int(\frac{0.75mc}{d(\text{Loc}(Q_0), \text{Loc}(Q_1)) + d(\text{Loc}(Q_1), \text{Loc}(Q_0))})$. Assuming $W$ workdays in a year, the probability of occurrence for any \emph{transitional} state $Q_i$ is then calculated as $P(Q_i) = (W / f_i) / W$.

The transition probability between states $Q_i$ and $Q_j$, denoted by $\chi_{i,j}$, is equivalent to the compound probability of the two independent states, i.e., $P(\chi_{i,j}) = P(Q_i)P(Q_j)$. The following conditions determine if a state $Q_i$ can transition to state $Q_j$: (1) $Q_i$ is a \emph{significant} state and the originating state for $Q_j$, (2) $Q_j$ is a \emph{significant} state and the destination state for $Q_i$, or (3) the two states originate from the same \emph{significant} state $Q_{s}$ and distance $d(\text{Loc}(Q_s),\text{Loc}(Q_i)) < d(\text{Loc}(Q_s),\text{Loc}(Q_j))$. The \emph{significant} states are always connected and their probabilities are calculated as $P(\chi_{0,1}) = 1 - \sum_{i=2}^{S-1} P(\chi_{0,i})$ and $P(\chi_{1,0}) = 1 - \sum_{i=2}^{S-1} P(\chi_{1,i})$, respectively.  All other transitions have a probability of 0.

\begin{figure}[t]
	\centering
	\includegraphics[width=.95\linewidth]{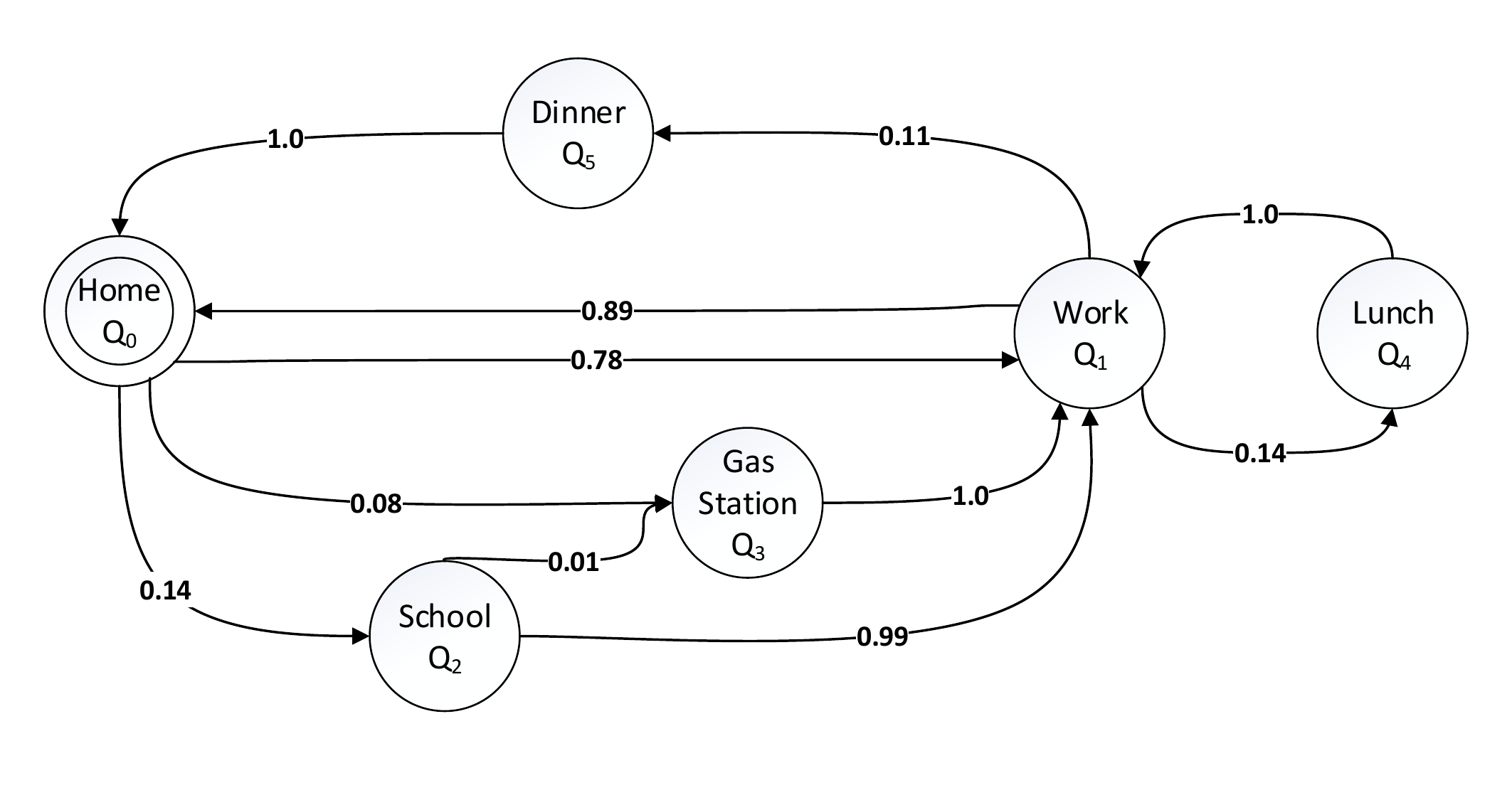}
	\caption{Example of a simplified finite state machine
          simulating a user's movements based on some transition
          probabilities.}
	\label{fig_statemachine}
\end{figure}

Note that users can go for `Lunch' in the afternoon and `Dinner' in the evening from the `Work' state. If we use the same `Work' state for both transitions, the probabilities are split when they clearly are different transitions. To address this, the `Work' state is internally represented as two states: $Q_{1a}$ for afternoon and $Q_{1e}$ for evening. Also note that the model described here is for weekdays, and a similar model is created for weekends with a different set of states (e.g., the user may leave from `Home' to watch a `Movie', eat `Dinner' and return `Home').

\Cref{fig_statemachine} provides an intuition for our automated finite state machine model. This specific model comprises of 6 states $Q = \{Q_0,\cdots, Q_5\}$ and their transition probabilities are shown. We see that it is possible to transition from state $Q_0$ to states $Q_1$, $Q_2$ or $Q_3$. As the transition probability $P(\chi_{0,1})$ is $0.78$, the model should typically choose state $Q_1$ $\approx8$ times out of $10$. This makes sense as a user will mostly go to `Work' from `Home' but may sometimes need to drop their kids to `School' or fill up gas at a `Gas Station'. 
\subsection{Modeling Mobility Trajectories}\label{sec_schedule}

The finite state machine generated for each user is used to synthesize mobility trajectories for the user every day. This is a 3 step process: (1) synthesize the user states for the entire day, (2) synthesize the schedule to satisfy the time constraints, and (3) synthesize the trajectory based on the schedule.

\subheading{Synthesizing the user states} 
The state machine of a user is loaded every day to generate a route of the states the user will visit that day. This route always starts and ends at the initial state $Q_0$ (`Home`) and traverses through $Q_1$ (`Work`), i.e., $R = \lbrack Q_0, \ldots, Q_1, \ldots, Q_0 \rbrack$.  The first state $Q_0$ can transition to any connected state $Q_i$ based on the transition probabilities of $Q_0$. The state $Q_i$ can then transition to any of its connected state $Q_j$ based on the transition probabilities of $Q_i$, and so forth forming a chain that ends at the final state $Q_0$. Note that the construction technique of the state machine ensures that this route traverses through $Q_1$. Let $P(\chi_i) = \{P(\chi_{i,0}), \ldots, P(\chi_{i,S-1})\}$ denote the set of all transitional probabilities of state $Q_i$. To obtain the next state, the system first derives a random transitional probability from a uniform distribution $\mathcal{P} = \mathcal{U}(0, 1)$. This probability $\mathcal{P}$ is then compared with the cumulative probabilities of all transitions in $P(\chi_i)$. A state $Q_j$ is selected if $\mathcal{P}$ lies between the previous state's cumulative probability and its cumulative probability, i.e., $P(X \le \chi_{i,j-1}) < \mathcal{P} \le P(X \le \chi_{i,j})$.

\subheading{Synthesizing the schedule} 
A realistic schedule should satisfy the time constraints set for every state in a user's state machine, such as arriving at work between $8am$ and $9am$ or dropping children to school before $8$:$30am$. The schedule should also satisfy the amount of time spent in each state, such as working for at least $8 hrs$. The schedule should also account for the time spent in transitioning from one state to the next, such as driving for $0.5 hrs$ to get from home to work. All these constraints can be formulated as linear equalities or inequalities, therefore, defining the problem of scheduling as a Linear Program (LP). Let $t^a_i$ and $t^d_i$ be the arrival and departure times at/from state $Q_i$. The above constraints can be formulated as follows: arriving at state $Q_i$ between $8am$ and $9am$ is formulated as $8am < t^a_i \leq 9am$, specifying that the user works at least $8hrs$ is formulated as $t^d_{i+1} - t^a_{i} \geq 8.0$, and the time spent in transitioning from home to work is formulated as $t^a_{i+1} - t^d_{i} = 0.5$. Naturally, all the times are specified in UTC for consistency and bounded by the day's limits (i.e., $00$:$00$:$00$ - $23$:$59$:$59$).

This set of linear equality and inequality constraints define a \textit{convex polytope} of all the schedules satisfying the state constraints, and the transition time constraints between the states. Let $T = (t^a_1, t^d_1, \ldots, t^a_S, t^d_S)$ denote a vector of all the arrival and departure time instants for a route containing $S$ states. One simple way of finding a point on this polytope is by defining an objective function for the vector $T$ with random coefficients, i.e., $c = (c_1, \ldots, c_{S})$ where $c_i\in[-1,1]$. Let $t(\chi_{i,j})$ denote the total time spent in transitioning between two states $Q_i$ and $Q_j$. Also, recall that $t_{min,i}$ specifies the minimum time spent in state $Q_i$ and $a_{min,i}$, $a_{max,i}$ specify the time bounds of arrival at the state (cf. \Cref{sec_states}). Using above attributes, the LP is formally defined as:

{\footnotesize \vspace{-1.5em}
\begin{align*}
\mbox{Maximize}          & \qquad \sum_{i=1}^{S} ({c_i}{t^a_i} + {c_i}{t^d_i})  & \text{where } \, c_i\in[-1,1] \\
\mbox{Subject to:}       & \qquad a_{min,j} < t^a_j \leq a_{max,j}              & \text{for } \, j = 1, 2, \ldots, S \\
                         & \qquad t^d_{j+1} - t^a_{j} \geq t_{min,j}            & \text{for } \, j = 1, 2, \ldots, S-1 \\
                         & \qquad t^a_{j+1} - t^d_{j} = t(\chi_{j,j+1})& \text{for } \, j = 1, 2, \ldots, S-1
\end{align*}
}%

Solving this LP identifies a corner of the polytope but not a random element within it. If the coefficients of the objective function were repeated, the LP will output the same schedule. To address this, we compute a random point within the polytope by finding different corners of the polytope using random coefficients and then computing a random linear combination of these corners. More precisely, let $C = \{C_1, \ldots, C_N\}$ denote a set of $N$ corners of the polytope obtained using random coefficients, and let $r = \{r_1, \ldots, r_N\}$ denote a set of positive random numbers such that $\sum_{i=1}^{N} r_i = 1$. The random solution defining the user's schedule for that day is then calculated as ${Schedule} = \sum_{i=1}^{N} {r_i}{C_i}$.

Note that as synthesizing the schedule using LP requires pre-calculated transition times $t(\chi_{i,j})$, the system calculates this time using the `pessimistic' traffic model of \emph{Google Maps Directions API}. The departure time is chosen as the mean of the time constraints for the start state. This typically gives us a worst case transition time between two states and can be used for scheduling. Note that for synthesizing the final trajectory, the `best\_guess' traffic model is used which provides more accurate traffic representation.

\subheading{Synthesizing the route between two states} 
The route between two synthetic states is generated using a graph $G = (V, E)$ constructed for the area. The system uses the \emph{Dijkstra's} algorithm to find the fastest route between the states, using the length and speed limit information present in each vertex. The resulting route is split into multiple waypoints based on turns and stop signs (extracted from OpenStreetMap).  These waypoints are given as input to the \emph{Google Maps Directions API} to obtain historical traffic information about the route.  The departure time is specified based on the schedule generated for that day. The route obtained from the Google API consists of multiple steps and can be represented as $R = [r_1, \ldots, r_S]$, where $S$ denotes the number of steps. Each step $r_i$ is attributed with geographic and traffic related information $r_i = (\mathcal{B}, d_{step}, t_{step})_i$, where $\mathcal{B}$ is the list of geographic coordinates of this step, $d_{step}$ is the length of this step, and $t_{step}$ is the time to traverse this step. 

\begin{figure}[t]
    \centering
    \begin{subfigure}[b]{.49\linewidth}
        \centering
        \includegraphics[width=\linewidth]{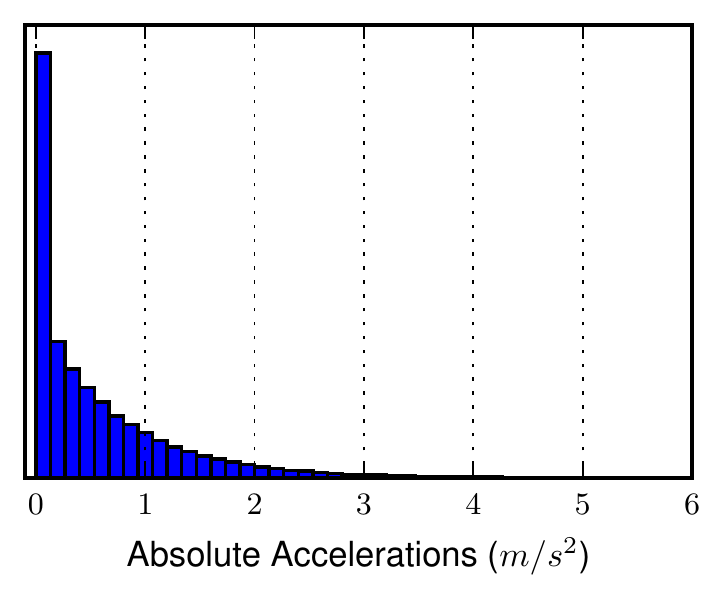}
        \caption{\label{fig_realaccs}Real Accelerations}    
    \end{subfigure}
    \begin{subfigure}[b]{0.49\linewidth}  
        \centering 
        \includegraphics[width=\linewidth]{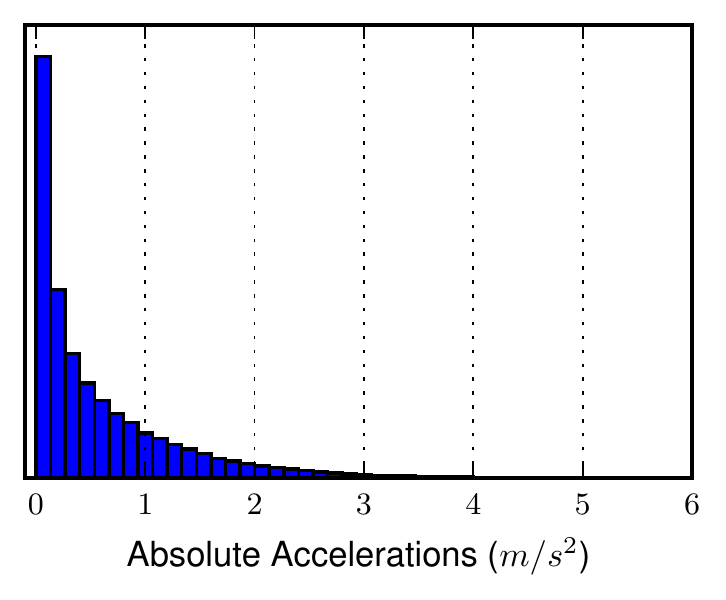}
        \caption{\label{fig_synaccs}Synthetic Accelerations}    
    \end{subfigure}
    \caption{Distribution of the absolute values of accelerations for both Real 
    ($\mu=0.61$, $M=0.34$, $\sigma=0.79$) and Synthetic ($\mu=0.61$, $M=0.32$, $\sigma=0.78$) routes.}
    \label{fig_hists}
\end{figure}

To generate realistic trajectories, all steps of a route must incorporate user driving behavior while also adhering to the step's traffic constraints, i.e., $d_{step}$ and $t_{step}$. To understand user driving behavior, we analyzed $400$ driving routes collected from 2 drivers and 4 phones (LG Nexus 5, LG Nexus 5X, Samsung Note 4, and Google Pixel). These routes covered a distance of $\approx 1400kms$ in a major city of USA consisting of both highway and internal roads, as well as peak and off-peak hours. The acceleration and speed information were extracted from these routes for every second to analyze their distribution. We found the speeds to be randomly distributed, however, the absolute values of accelerations approximate to an exponential distribution (mean $\mu=0.61$, median $M=0.34$, and standard deviation $\sigma=0.79$) shown in \Cref{fig_realaccs}. Note that the distribution is an approximation and not truly exponential because $\mu < \sigma$, where $\mu=\sigma$ is a property of exponential distributions. Analyzing individual routes, the range of means of absolute accelerations, denoted by $[\bar{|a|}_{min}, \bar{|a|}_{max}]$, varied between $0.1m/s^2$ and $1.1m/s^2$. The range of standard deviations of absolute accelerations, denoted by $[\sigma(|a|)_{min}, \sigma(|a|)_{max}]$, were between $0.4m/s^2$ and $1.1m/s^2$. The bounds of all acceleration values, denoted by $[a_{min}, a_{max}]$, were between $-7m/s^2$ and $7m/s^2$. The means of the accelerations were $\approx 0m/s^2$ for every route. 

The above constraints can be formulated as a list of equalities and inequalities, this time defining a non-linear constraint optimization problem. Such problems can be solved by using Sequential Quadratic Programming (SQP) methods. Let $a = (a_1, \ldots, a_N)$ denote a vector of acceleration values for each step, where $N$ denotes the travel time of the step, i.e., $N = int(t_{step})$. Let $v_0$ denote the initial speed coming into this step and $v = (v_1, \ldots, v_N)$ denote a vector of speeds calculated from $v_0$ and the vector $a$. The objective of this optimization is to find an optimal vector $a$ that minimizes $|\bar{v} - (d_{step}/t_{step})| < \Delta$ to adhere to the traffic constraints, where $\bar{v}$ is the mean of vector $v$, and $d_{step}$, $t_{step}$ represent the step's distance and time. The $\Delta$ is a threshold that determines whether the minimized objective function value is acceptable. All rejected optimizations are retried with a higher number of iterations till a valid solution satisfying the threshold is found. We observed that this optimization typically yields an optimal vector $a$ that approaches the lower mean bound of the absolute accelerations $\bar{|a|}_{min}$, for most optimizations. To address this, we derive a new lower mean bound for every route from a uniform distribution and use the following range for optimization: $[\bar{|a|}_{rand}, \bar{|a|}_{max}]$, where $\bar{|a|}_{rand} = \mathcal{U}(\bar{|a|}_{min}, \bar{|a|}_{max} - \delta)$, and $\delta$ is a small constant to ensure that $\bar{|a|}_{rand} < \bar{|a|}_{max}$. The optimal vectors $a_i$ for every step $i$ are merged to represent the route's accelerations. Note that a bounded constraint of the form $x_1 \le x \le x_2$ can be rewritten as $(x_2 - x)(x - x_1) \ge 0$ for simplifying the constraint for the solver. Using above attributes, the route optimization for each step is formally defined as:

{\footnotesize  \vspace{-1.5em}
\begin{align*}
\mbox{Minimize}           & \qquad |\bar{v} - (d_{step}/t_{step})| \\
\mbox{Subject to:}        & \qquad \bar{a} = 0 \\
                          & \qquad (\bar{|a|}_{max} - \bar{|a|})(\bar{|a|} - \bar{|a|}_{rand}) \ge 0 \\
                          & \qquad (\sigma(|a|)_{max} - \sigma(|a|))(\sigma(|a|) - \sigma(|a|)_{min}) \ge 0 \\
                          & \qquad \sigma(|a|) - \bar{|a|} \ge 0 \\
\mbox{Bounds:}            & \qquad a_{min} \le a_j \le a_{max} \hspace{25pt} \text{for } \, j = 1, 2, \ldots, N
\end{align*}
}%

Some additional constraints applied to the optimization are that $v_0=0$ for the first step and $v_N=0$ for the last step of the route. The optimization is improved by providing an initial guess of bounded accelerations from a gaussian distribution $\mathcal{N}(\bar{v'}, 2)$, where $\mu=\bar{v'}$ is the mean step speed, i.e., $\bar{v'} = d_{step}/t_{step}$, and $\sigma=2m/s$ is the standard deviation of the speed. \Cref{fig_synaccs} shows the distribution of the absolute accelerations generated for synthetic trajectories. We can observe that the parameters and shape of the distribution closely follows the parameters and shape of the real distribution. 

Note that this work uses a linear model for synthesizing walks from a state's coordinates to a graph vertex, and vice versa. The vertex containing a point nearest to the state's coordinates is chosen, and the driving route is started/stopped at this point. This simple model assumes a constant walking speed as our main focus was on driving. We plan to study models for generating realistic walk patterns in the future. Also note that as GPS accuracy varies, a small random gaussian noise is added to each coordinate of the final trajectory.

\section{Evaluation}
\label{sec_mitigation_evaluation}

In this section, we evaluate \system~using the following metrics: the portability, stability and performance of the system, and the detection of synthetic trajectories by popular location-driven apps, by regular users, and by Machine Learning algorithms.

\subsection{System Portability and Stability}
\system~is compatible with Android \code{KitKat} and onwards. It has been tested to work on Xposed Framework API versions 82 to 89 (current) which are compatible with the above Android versions. This implies that \system~can be ported to $\approx94\%$ of all Android devices globally (based on information from the Android Dashboard~\cite{android_dashboard} as of August 10, 2018).

The system's stability was evaluated on 4 smartphones and the results are shown in \Cref{tab_stability}. The evaluation was performed using $1000$ popular apps on Google Play Store that requested location permissions or accessed the sensors. All the apps had a minimum rating of $4.0$ and a minimum vote count of $10,000$ users. These $1000$ apps were successively run twice using an automated UI application exerciser tool called Android Monkey \cite{monkey}, once on a stock Android version of these smartphones and then with \system~installed on the same phones. The tool was configured to stress test each app's activities to monitor how many additional apps crash or fail to execute. The same settings were used for both tests ($seed=1$, $num\_events=2500$) to ensure that the same pseudo-random events were generated.

The first row for each phone in \Cref{tab_stability} shows the test results for the stock version and the second row shows the test results for \system. All the apps installed and ran on every phone except for 15 apps on the HTC One M9 (possibly due to compatibility reasons). The number of successful monkey runs are very similar in both the tests with the stock version performing better on two phones and the \system~version performing better on the other two. We analyzed the errors/crashes manually to check for Xposed or \system~specific errors and did not find any. This validates that \system~remains stable and runs as expected for different devices, OS versions, apps and in heavy use.

\begin{table}[t]
\centering
\caption{Results of the Stability evaluation for \system~using $1000$ popular Android apps on $4$ smartphones.\label{tab_stability}}
\begin{tabular}{ p{1.75cm} p{1.75cm} l l l }
	\hline
	\textbf{Phone}                  & \textbf{Version}               & \textbf{Installed}   & \textbf{Success}   & \textbf{Failure} \\
	\hline 
	\multirow{2}{4em}{HTC~One~M7}   & \multirow{2}{4em}{Lollipop}    & 1000                     & 892                & 108 \\  
	                                &                                & 1000                     & 894                & 106 \\ 
	\hline  
	\multirow{2}{4em}{HTC~One~M9}   & \multirow{2}{4em}{Marshmallow} & 985                    & 796                & 189 \\  
	                                &                                & 985                    & 791                & 194 \\ 
	\hline  
	\multirow{2}{4em}{LG~Nexus~5}   & \multirow{2}{4em}{Lollipop}    & 1000                     & 938                & 62 \\  
	                                &                                & 1000                     & 944                & 56 \\ 
	\hline  
	\multirow{2}{4em}{LG~Nexus~5X}  & \multirow{2}{4em}{Marshmallow} & 1000                     & 851                & 149 \\  
	                                &                                & 1000                     & 848                & 152 \\
	\hline  
\end{tabular}
\end{table}

\subsection{System Performance}
\label{sec_performance}

\system~was extensively evaluated for performance overheads occurring from the most expensive operations of the system. We identified 3 potential performance bottlenecks in our system: (1) the API call interception function using the Xposed framework; (2) the add audit event function of the PrivoScope service; and (3) the location provider function of the Synthetic Location service. We implemented a test app that invoked these functions $1$ million times to test performance.  The execution time was calculated as the difference between two \code{System.nanoTime} method calls placed immediately before and after the function execution. The API interception bottleneck is caused by the Xposed framework loading and hooking method calls. To evaluate its performance, we created an empty method inside our system and hooked it using the Xposed framework.

\Cref{tab_performance} shows the mean $\mu$, standard deviation $\sigma$ and maximum time of execution for the three functions on a LG Nexus 5 and a LG Nexus 5X. The API interception function using the Xposed framework averaged about $\mu=0.2ms$ on both the phones, which is negligible from a usage perspective. The add audit event function of \privo~had a low $\mu$ for both the phones ($4.3ms$ and $3.2ms$, resp), and its performance is also acceptable. The location provider function of the Synthetic Location service had a relatively higher $\mu$ and $\sigma$ for the Nexus 5 ($\mu=11.1ms$, $\sigma=7.7ms$). We believe this overhead is due to database lookups performed by the service to check the location preferences for the app. Overall, the entire system can run with an average overhead of $15.6ms$ on the Nexus 5 and $9.1ms$ on the Nexus 5X which should have a negligible impact on the user experience. The sum of worst case performances overhead at $171.8ms$ on the Nexus 5 should also not affect user experience since such overhead occurs rarely.

\begin{table}[t]
\centering
\caption{Results of the Performance evaluation of \system~for $2$ smartphones.\label{tab_performance}}
\begin{tabular}{ l p{2.2cm} l l l }
	\hline
	\textbf{Phone}                      & \textbf{Service}     & \textbf{Mean ($\mu$)}   & \textbf{Std ($\sigma$)}    & \textbf{Max} \\
	\hline
	\multirow{3}{4em}{Nexus~5}          & Xposed Hook     & 0.2 ms                  & 0.3 ms                       & 17.1 ms \\
	                                    & Add Audit Event      & 4.3 ms                  & 3.8 ms                       & 67.1 ms \\
	                                    & Update Location      & 11.1 ms                 & 7.7 ms                       & 87.6 ms \\
	\hline
	\multirow{3}{4em}{Nexus~5X}         & Xposed Hook     & 0.2 ms                  & 0.15 ms                      & 5.7 ms \\
	                                    & Add Audit Event      & 3.2 ms                  & 1.6 ms                       & 26.8 ms \\
	                                    & Update Location      & 5.7 ms                  & 1.5 ms                       & 16.0 ms \\ 
	\hline
\end{tabular}
\end{table}

\subsection{Detection of Synthetic Trajectories}

\subsubsection{Detection by Popular Mobile Apps}

\begin{table*}[ht]
\centering
\caption{Results of the Synthetic Trajectories detection test on $10$ popular Android apps that rely on location data.\label{tab_detection}}
\begin{tabular}{ p{2.5cm} p{2.5cm} p{1.2cm} p{2.5cm} p{2.5cm} p{3cm} }
	\hline
	\textbf{App Name}      & \textbf{Category} 	& \textbf{Rating} & \textbf{Synthetic}   & \textbf{High Speed (HS)}     & \textbf{HS+Teleport (HS+T)} \\
	\hline 
	Ingress                & Adventure	Game		& 4.3				& \checkmark        & Detected            & Detected \\  
	Pok\'emon Go           & Adventure	Game	 	& 4.1 				& \checkmark        & \checkmark          & \checkmark \\  
	Geocaching             & Health \& Fitness		& 4.0 				& \checkmark        & \checkmark          & \checkmark \\  
	Glympse                & Social			& 4.5				& \checkmark        & \checkmark          & \checkmark \\  
	Family Locator         & Lifestyle			& 4.4 				& \checkmark        & \checkmark          & \checkmark \\  
	happn                  & Lifestyle			& 4.5				& \checkmark        & \checkmark          & \checkmark \\  
	Yelp                   & Travel \& Local		& 4.3 				& \checkmark        & \checkmark          & \checkmark \\  
	Foursquare             & Food \& Drink			& 4.1				& \checkmark        & \checkmark          & \checkmark \\  
	Waze                   & Maps \& Navigation	& 4.6 				& \checkmark        & \checkmark          & Unstable \\  
	Google Maps            & Travel \& Local 		& 4.3 				& \checkmark        & \checkmark          & Unstable \\   
	\hline
\end{tabular}
\end{table*}

We evaluated this metric using 10 popular location-driven apps (listed in \Cref{tab_detection}) on Google Play Store. These apps rely heavily on location data to provide their services to users. The evaluation was performed by feeding these apps three types of synthetic location data and monitoring their behavior. In test 1 \textbf{(Synthetic)}, the synthetic trajectories were generated using the techniques described in \Cref{sec_schedule}. In test 2 \textbf{(HS)}, the trajectories from test 1 were time compressed by a factor of 5 such that the user appeared to move 5 times faster (e.g., at $300km/h$ in a $60km/h$ speed zone). In test 3 \textbf{(HS+T)}, the trajectories from test 2 were perturbed by large noises ($\approx 1000m$) such that the user appeared to teleport to different locations very quickly. The expected results was that apps that detect fake location should be able to easily detect the \textbf{HS} and \textbf{HS+T} trajectories, but not \textbf{Synthetic} trajectories.  

\Cref{tab_detection} shows the results of the three tests for our test apps. None of the apps were able to detect synthetic locations in the \textbf{Synthetic} trajectories test. Even for \textbf{HS} and \textbf{HS+T} trajectories test, with the exception of Ingress, none of the other apps detected the presence of high speed and noisy synthetic locations. Ingress did not ban us from playing the game, however, it denied points when it detected that the user was moving too fast or teleporting. Pok\'emon Go is also known to ban users, however, we did not get banned during our tests even after capturing many Pok\'emons using the noisy data. This is likely because the ban threshold is set to high to prevent users from going to a higher level by cheating. All the remaining apps kept performing their functions without detecting the presence of the synthetic data. Note that Waze and Google Maps navigation operated properly for \textbf{HS} but became unstable for \textbf{HS+T}, which was expected as they constantly updated the\ routing information based on the teleported locations. 

These observations indicate that popular location-driven apps fail to check validity of the received data. Some of these apps (Ingress, Pok\'emon Go, Foursquare and Google Maps) check whether the \code{MockLocationProvider} \cite{mocklocation} is enabled on the device. Some apps rely on other schemes to limit user abuse (e.g., Foursquare detects and limits rapid check-ins). This means that they rely on simple checks but do not implement algorithms for detecting synthetic data. The only app that checked location validity in our set was Ingress, and it was unable to detect any discrepancies in the synthetic trajectories generated by our system. 

\subsubsection{Detection by Regular Users}

\begin{figure}[t]
    \centering
    \begin{subfigure}{\linewidth}
        \centering
        \fbox{\includegraphics[width=\linewidth]{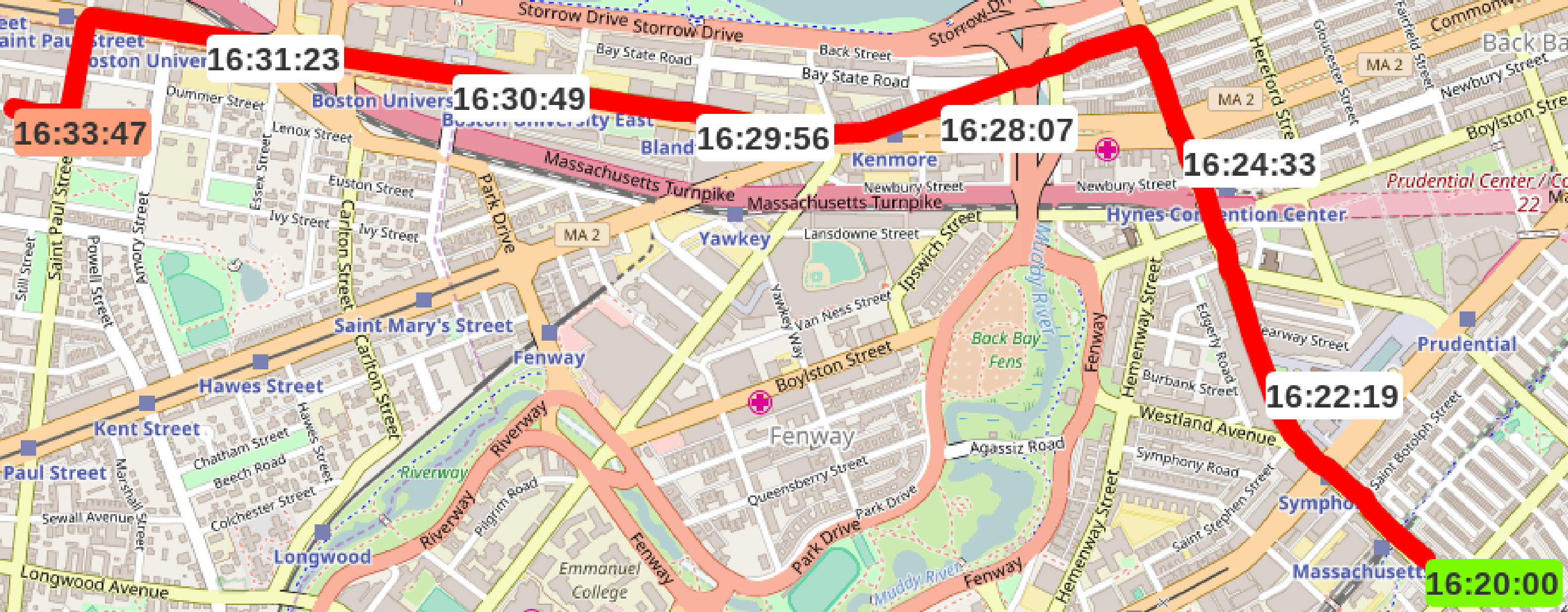}}
        \caption{\label{fig_real} Real Driving Route}
    \end{subfigure}
    \begin{subfigure}{\linewidth}
        \centering
        \fbox{\includegraphics[width=\linewidth]{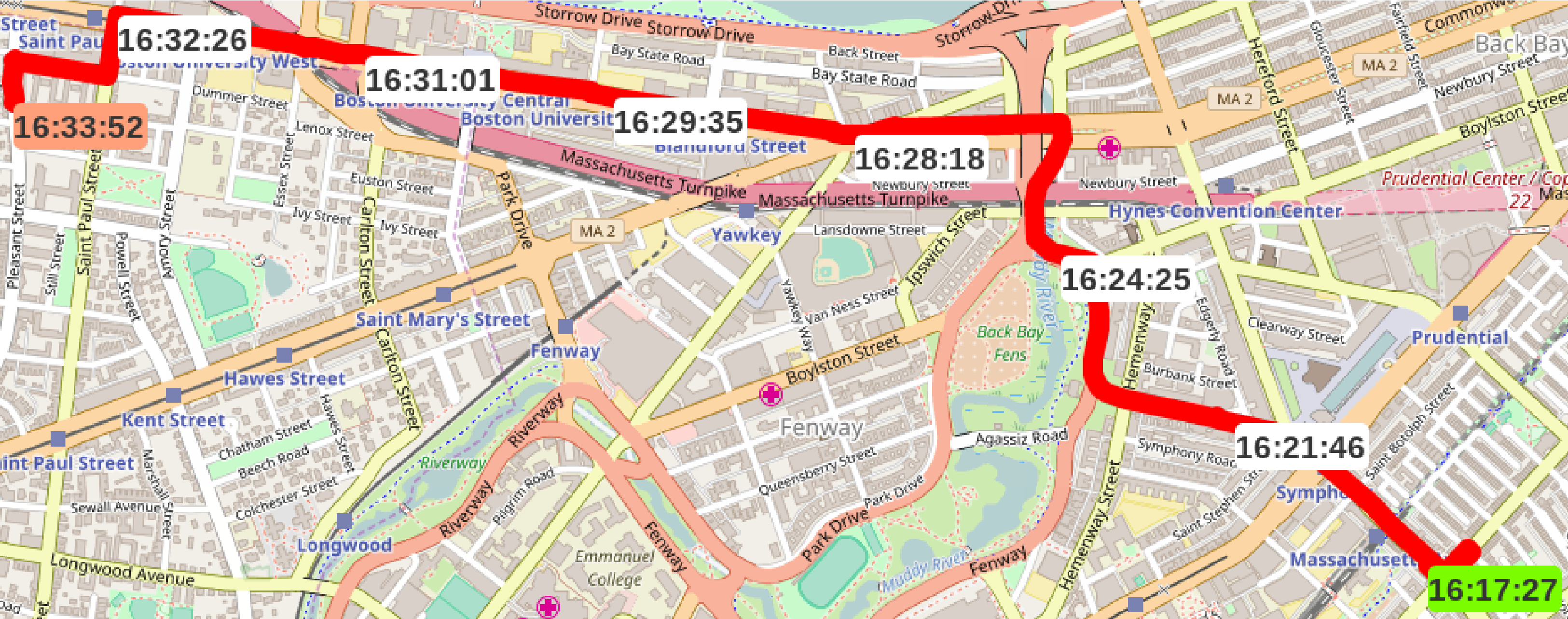}}
        \caption{\label{fig_synthetic} Generated Synthetic Trajectory}
    \end{subfigure}

    \caption{An example of the similarity between a real route and a generated synthetic route.}
    \label{fig_realvssynthetic}
\end{figure}

\begin{table}[t]
\centering
\caption{Cumulative results of the User Study on Mechanical Turk sorted by the number of noisy trajectories correctly labeled.\label{tab_cum_turk}}
\begin{tabular}{ l l c c c c }
    \hline
    \textbf{Noisy}          & \textbf{Surveyors} & \multicolumn{2}{c}{\textbf{Real Trajectories}}    & \multicolumn{2}{c}{\textbf{Synthetic Trajectories}} \\
                            &                    & \textbf{Real}           & \textbf{Synthetic}      & \textbf{Real}           & \textbf{Synthetic}\\
    \hline 
    0                       & 100                & 65.1\%                  & 34.9\%                  & 66.0\%                  & 34.0\%\\ 
    1                       & 91                 & 65.4\%                  & 34.6\%                  & 65.9\%                  & 34.1\%\\ 
    2                       & 72                 & 65.7\%                  & 34.3\%                  & 65.4\%                  & 34.6\%\\ 
    3                       & 54                 & 68.3\%                  & 31.7\%                  & 64.4\%                  & 35.6\%\\ 
    \hline
\end{tabular}
\end{table}

We evaluated this metric by conducting two separate user studies: one comprising of a group of $12$ students from a university and another comprising of $100$ users from Amazon Mechanical Turk \cite{turk}. The intuition behind two studies was to understand the results from two perspectives; one of users who know the area very well and another of users unaware of the area. The university area was chosen so that the students were aware of its traffic congestions. The study asked the users to visually analyze a mix of $20$ real and synthetic trajectories and label them as `Real' or `Synthetic' based on their observations. \Cref{fig_real,,fig_synthetic} show an example of a real route and a synthetic trajectory used for the study. The green marker marks the start location, the white markers are $500 m$ apart, and the red marker marks the stop location. These markers display the time the vehicle  was at the given location.

The trajectories were created as follows: First, we drove $10$ unique routes close to the university area, each starting and ending at different locations and  times of the day. Each route can be represented as $R = [n_1, \ldots, n_L]$, where $n$ is a node and $L$ is the number of nodes in the route. Each node $n_i$ is attributed with timing and geographic information $n_i = (t_i, \text{Loc}(n_i))$, where $t_i$ is the timestamp and $\text{Loc}(n_i)$ is the node's geographic coordinates. Next, we generated $10$ synthetic routes similar to the $10$ real routes using the timestamp of the first node (i.e., $t_1$) and geographic coordinates of the end nodes (i.e., $\text{Loc}(n_1)$ and $\text{Loc}(n_L)$) for each route R. The trajectories were shuffled so they appeared in a random order. For mechanical turk, we added three very noisy trajectories which looked obviously synthetic to find users who did not take the study seriously.

\subheading{University Students Study}
For the real trajectories, $\approx64.2\%$ of the trajectories were labeled as `Real' and the rest were labeled as `Synthetic'. For the synthetic trajectories, $\approx 65.8\%$ of the trajectories were labeled as `Real' and the rest were labeled as `Synthetic'. Note that more users of this study confused the `Synthetic' trajectories to be `Real'.

\subheading{Amazon Mechanical Turk Study} 
For the real trajectories, $\approx 68.3\%$ of the trajectories were labeled as `Real' and the rest were labeled as `Synthetic'. For the synthetic trajectories, $\approx 64.4\%$ of the trajectories were labeled as `Real'. The above results are for $54$ users who detected all the obviously noisy trajectories. \Cref{tab_cum_turk} shows the cumulative results of the mechanical turk study based on the number of noisy trajectories detected by the users. We can see that the results are not significantly different even for all 100 users, however, more users labeled `Synthetic' as `Real'.

The results indicate that it was difficult for the users to differentiate between synthetic and real driving trajectories. There was confusion in both groups regarding their validity. Evaluating individual trajectories, we saw that this confusion applied to each trajectory as not a single one was labeled as `Real' or `Synthetic' unanimously by all users.

\subsubsection{Detection by Machine Learning Algorithms}

We evaluated this metric using the $400$ routes collected for analyzing user driving behavior (cf. \Cref{sec_schedule}). These set of routes were labeled as `Real'. For each real route, a corresponding synthetic route was generated using the real route's departure time, and start and end locations. These set of routes were labeled as `Synthetic'. Note that the routes may have completely different trajectories, therefore, extracting spatial features or using the entire route for machine learning will generate inaccurate models. As such, we extract only temporal features from these routes. The following 9 features were extracted from both set of routes for training the machine learning models: \emph{max and min acceleration}, \emph{mean and standard deviation of accelerations}, \emph{mean and standard deviation of absolute accelerations}, \emph{maximum speed}, \emph{idle time} and \emph{distance traveled}. The models were built and the predictions were averaged over $1000$ iterations. In each iteration, 90\% of the dataset from each set were randomly chosen for training data, and the remaining 10\% from each set were test data.

\Cref{tab_mlresults} shows the list of algorithms that were evaluated and their prediction accuracies for the `Real' and `Synthetic' test trajectories. Note that in our context, the ideal results should be a 50-50 split, i.e., 50\% of `Real' routes are predicted as `Synthetic' and 50\% of `Synthetic' routes are predicted as `Real'. We can observe that most algorithms (except Decision Trees) have an average prediction accuracy of close to 50\%. Three of those algorithms (Naive Bayes, Neural Network and SVM) display results biased towards one of the two classifiers implying that the models had difficulty predicting the correct classifier and defaulted to one classifier. The Decision Trees algorithm could detect $\approx62\%$ of the `Synthetic' trajectories as synthetic. The ensemble classifier of Decision Trees, Random Forest, could detect $\approx63\%$ of the `Synthetic' trajectories as synthetic. These numbers also do not signify large detection rate for our synthetic trajectories. We must note that this evaluation is preliminary as $800$ routes do not suffice for these algorithms to build generalized models from training data, and the models may be subject to overfitting. We intend to extend our dataset in the future to incorporate more routes and run this evaluation again for more generalized models. 

\begin{table}[t]
\centering
\caption{Results of the Machine Learning algorithms evaluation showing the `Real' and `Synthetic' prediction accuracy.\label{tab_mlresults}}
\begin{tabular}{ l c c c c }
	\hline
	\textbf{Algorithm}      & \multicolumn{2}{c}{\textbf{Real Trajectories}}			& \multicolumn{2}{c}{\textbf{Synthetic Trajectories}}		\\
	 						& \textbf{Real} & \textbf{Synthetic}						& \textbf{Real} & \textbf{Synthetic}				      	\\
	\hline  
	Decision Trees			& 53\%			& 47\%										& 38\%			& 62\%										\\
	Random Forest			& 61\%			& 39\%										& 37\%			& 63\%										\\
	Nearest Neighbor		& 50\%			& 50\%										& 43\%			& 57\%										\\
	10 Nearest Neighbor		& 49\%			& 51\%										& 43\%			& 57\%										\\
	Naive Bayes				& 86\%			& 14\%										& 86\%			& 14\%										\\
	Neural Networks			& 95\%			& 5\%										& 96\%			& 4\%										\\
	SVM						& 5\%			& 95\%										& 3\%			& 97\%										\\			
	\hline
\end{tabular}
\end{table}

\section{Related Work}
\label{sec_mitigation_relatedwork}

A large body of research has focused on mitigating location and other private information leakage attacks on Android devices. Most of these works are orthogonal to our system as their motivation and techniques differ. Examples of such work include, but are not limited to, recommending new frameworks/privacy metrics \cite{Fawaz:2014:LPP:2660267.2660270, boxify, spe, spism, deva2015, fawazu2015, OyaTP2017}, location obfuscation \cite{Andres2013, Bordenabe2014, Shokri2012, Wang2012, ardagna2011}, location cloaking \cite{Hoh07}, generating dummy locations \cite{Kato2012, Kido2005, Lu2008, Suzuki2010, You2007, fakemask}, tainting sensitive data \cite{taintdroid, taintart}, dynamic analysis \cite{appaudit, tissa}, static code analysis \cite{flowdroid, amandroid, rdroid, lu_ndss15}, permissions analysis \cite{drandroid}, and application retrofitting \cite{retroskeleton, appfence, Case16}.

Synthesizing human mobility has also been studied in the context of opportunistic networks \cite{orbit, slaw, ekman2008, karamshuk2011}, ad-hoc and vehicular wireless networks \cite{tuduce2005, choffnes2005, kim2006, zheng2010}, community based mobility models \cite{hong1999, li2002, hermann2003, musolesi2006}, predicting location of moving objects \cite{tao2004, zhou2007}, and implementing efficient location update mechanisms \cite{wolfson2003, ding2009, huang2013, huang2014}. Some research has also focused on generating synthetic traces for user privacy \cite{Bindschaedler2016, krumm2009, chow2009}, however, these works have limitations that can enable an adversary to detect fake traces. None of the above works satisfy traffic constraints for different roads at different times of the day, nor take into account the statistical properties of user driving behavior. For example, \cite{krumm2009, wolfson2003} simply superimpose speed patterns from real routes on synthetic traces based on the street type without accounting for traffic conditions of the road. These speed patterns can also be repeated and can be detected. Bindschaedler and Shokri~\cite{Bindschaedler2016} generate synthetic traces that are derived from seed datasets of real traces which does not apply in our context of generating completely synthetic identities for users. Their work is also not scalable globally due to their reliance of real datasets, while ours can easily scale since map data for any location of the world is readily available.

Beresford et al.~\cite{mockdroid} implemented MockDroid, a modified version of Android 2.2.1 with a user controlled permissions manager. The system allowed users to define mock permissions for installed apps. The location mock permission was implemented to block all location fixes from reaching the app simulating a lack of available location information. The authors ran the system on 27 apps and showed that most apps continued to function with reduced functionality. This work is similar to the current Android permissions model and, therefore, subject to the weaknesses in Android's permission model that we have addressed with \system. 

Agarwal and Hall~\cite{pmp} implemented ProtectMyPrivacy (PMP) for iOS jailbroken devices that intercepted method calls accessing user's private data, and allowed the user to substitute anonymized data in place of the real information. The limitations of this work are: (1) the anonymized data is provided by the user at run-time which may be completely random and unrealistic, and (2) the app's functionality is paused for user input which is detrimental to user experience and possibly also to the app's functionality. The above limitations are addressed by \system~through the seamless delivery of realistic synthetic locations to apps without requiring constant interaction with the user.

Liu et al.~\cite{ppa} implemented Personalized Privacy Assistant (PPA) for rooted Android devices. This system is a modified \code{App Ops} permission manager that displays an app's recent requests and the frequencies of requests in the past 7 days. The system uses this information to generate daily privacy nudges to motivate users to interact and change their privacy settings. We believe that this information provided to users is not sufficient to make any kind of privacy-aware decisions. Additional context is required to determine if an app is misusing the information (e.g., time and duration of those requests, was the app in the background?). \privo~addresses the limitations by providing much more context to the users and displaying them in a way that it is easier for the users to grasp and visualize these accesses.

Zheng et al.~\cite{zheng2010} propose an agenda driven mobility model that considers a person's daily social activities for motion generation.  They derive this agenda from the National Household Travel Survey (NHTS) database by the U.S. Department of Transportation. The first agenda and all subsequent activities are based on the NHTS activity distribution, and addresses are picked at random from many addresses for the corresponding activity. The start time of the first agenda determines the schedule for the entire day and each activity starts immediately after the mean dwell time+longest transition time from previous activity. The route between two activities assumes a longest possible time given by the Dijkstra's algorithm. This work has several limitations (all addressed by \system) that are trivial to detect: (1) the addresses are picked at random without accounting for distances (e.g., gas station may be miles away from regular route), (2) the routes do not incorporate any traffic information and are always static, and (3) the routes do not incorporate any driving behavior and likely assume a constant speed of motion.

Fawaz and Shin~\cite{Fawaz:2014:LPP:2660267.2660270} implemented LP-Guardian, a privacy protection framework modifying the Android source code. The framework changes location granularity of installed apps based on the threat posed by the app and its location granularity requirements. It automatically coarsens the location to a city level if it identifies a request from an A\&A library, the app is in the background, or the app is a weather app. It synthesizes the location for fitness apps but preserves features of the actual route such as the distance traveled. The framework supplies a synthetic location if it determines that it is not safe to release the location. This work has the following limitations that are addressed in \system: (1) the preservation of route features can lead to inference of the user's real locations, and (2) unless chosen very carefully, the synthetic traces generated from real features will not snap to streets (e.g., different street lengths and curvatures) and can be detected as synthetic.

% \begin{table*}[t]
% \centering
% \caption{Summary of the protections implemented by current privacy protection systems in comparison with \system.}
% \begin{tabular}{ p{3.8cm} p{1.7cm} p{1.7cm} p{1.7cm} p{1.7cm} p{1.8cm} p{1.7cm} }
%   & \textbf{MockDroid} & \textbf{PMP} & \textbf{PPA} & \textbf{Zheng et al.} & \textbf{Fawaz \& Shin} &\textbf{\system} \\
%   \hline
%   \textit{System Implemented} & \checkmark & \checkmark & \checkmark & & \checkmark & \checkmark \\
%   \textit{Addresses Weakness 1 (W1)} & \checkmark & & \checkmark & & & \checkmark \\
%   \textit{Addresses Weakness 2 (W2)} & & & & & & \checkmark \\
%   \textit{Addresses Weakness 3 (W3)} & & & \checkmark & & & \checkmark \\
%   \textit{Addresses Weakness 4 (W4)} & & \checkmark & & & \checkmark & \checkmark \\
%   \textit{Addresses Weakness 5 (W5)} & & & & \checkmark & \checkmark & \checkmark \\
% \end{tabular}
% \label{tab_comparison}
% \end{table*}

\section{Conclusion}
\label{sec_mitigation_conclusion}

We presented \system, a system that addresses current privacy protection weaknesses in Android, provides users with a tool to analyze how apps access their private information, and the capability to provide obfuscated/synthetic data to untrusted apps. We demonstrated that~\system~is portable to most Android devices globally, is reliable, has low-overhead, and generates privacy-preserving synthetic trajectories that are difficult to differentiate from real mobility trajectories by an adversary.

\bibliographystyle{IEEEtran}
\bibliography{paper}

\end{document}